\documentclass{aa}  
\usepackage{graphicx}
\usepackage{txfonts}
\newcommand{\ha}{H$\alpha$}

\newcommand{\htwo}{H$_2$}
\newcommand{\hi}{\ion{H}{i}}

\newcommand{\fei}{\ion{Fe}{i}}
\newcommand{\feii}{[\ion{Fe}{ii}]}

\newcommand{\hei}{\ion{He}{i}}

\newcommand{\av}{$A_V$}
\newcommand{\kms}{km\,s$^{-1}$}
\newcommand{\um}{$\mu$m}

\newcommand{\lsun}{L$_{\odot}$}
\newcommand{\msun}{M$_{\odot}$}
\newcommand{\msunyr}{M$_{\odot}$\,yr$^{-1}$}
\newcommand{\macc}{$\dot{M}_{acc}$}

\newcommand{\lacc}{$L_{\mathrm{acc}}$}

\begin{document}

   \title{High-resolution TNG spectra of T Tauri stars:}
     \subtitle{Near-IR GIANO observations of the young variables XZ Tau and DR Tau}

   \author{S. Antoniucci\inst{1}
          \and B. Nisini\inst{1}
          \and K. Biazzo\inst{2}
          \and T. Giannini\inst{1}
          \and D. Lorenzetti\inst{1}     
          \and N. Sanna\inst{3} 
          \and A. Harutyunyan\inst{4} 
          \and L. Origlia\inst{5} 
          \and E. Oliva\inst{3}
          }

   \institute{INAF - Osservatorio Astronomico di Roma, Via di Frascati 33, 00040 Monte Porzio Catone, Italy
        \and  INAF-Osservatorio Astrofisico di Catania, via S. Sofia 78, 95123, Catania, Italy
        \and INAF-Osservatorio Astrofisico di Arcetri, Largo Enrico Fermi 5, I-50125 Firenze, Italy
                \and Telescopio Nazionale Galileo, Fundación Galileo Galilei - INAF, Rambla José Ana Fernández Pérez, 7, E-38712 Breña Baja, TF, Spain
                \and INAF-Osservatorio Astronomico di Bologna, via Ranzani 1, 40127, Bologna, Italy
              }

    \date{}
        \titlerunning{GIANO high-resolution near-IR spectra of the young variables XZ Tau and DR Tau}
        \authorrunning{S. Antoniucci et al.}

  \abstract
{}
{
We aim to characterise the star-disk interaction region in T Tauri stars that show photometric and spectroscopic variability.
}
{
We used the GIANO instrument at the Telescopio Nazionale Galileo to obtain near-infrared high-resolution spectra
(R$\sim$ 50,000) of XZ Tau and DR Tau, which are two actively accreting T Tauri stars classified as EXors. 
Equivalent widths and profiles of the observed features are used to derive 
information on the properties of the inner disk, the accretion columns, and the winds. 
}
{
Both sources display composite \hi\ line profiles, where contributions from both accreting gas and high-velocity winds can be recognised. These lines are progressively more symmetric and narrower with increasing upper energy
which may be interpreted in terms of two components with different decrements or imputed to self-absorption effects.
XZ Tau is observed in a relatively high state of activity with respect to literature observations.
The variation of the \hei\ 1.08\um\, line blue-shifted absorption, in particular,  suggests that
the inner wind has undergone a dramatic change in its velocity structure, connected
with a recent accretion event.
DR Tau has a more stable wind as its \hei\ 1.08\um\, absorption does not show variations with time in spite of strong variability of the emission component. 
The IR veiling in the two sources can be interpreted as due to blackbody emission
at temperatures of 1600 K and 2300 K for XZ Tau and DR Tau, respectively, with emitting areas
$\sim$ 30 times larger than the central star. While for XZ Tau these conditions are consistent with emission from
the inner rim of the dusty disk, the fairly high temperature 
inferred for DR Tau might suggest that its veiling originates from a thick gaseous disk located within
the dust sublimation radius.
Strong and broad metallic lines, mainly from \ion{C}{i} and \fei, are detected in XZ Tau, similar to those 
observed in other EXor sources during burst phases. At variance, DR Tau shows weaker and narrower metallic lines, 
despite its larger accretion luminosity. This suggests that accretion is not the only driver of metallic line
excitation. 
}
{
The presented observations demonstrate the potential of wide-band, high-resolution 
near-IR spectroscopy to simultaneously probe the different phenomena that occur in the interaction region
between the stellar magnetosphere and the accretion disk, thus providing hints on how these two structures are
linked to each other.
}


\maketitle

\section{Introduction}
\label{sec:intro}

Accretion, jets, and winds are phenomena that characterise the pre-main sequence phase of stellar 
evolution (Classical T Tauri, or CTT, phase). 
They are also the main mechanisms causing gas 
dispersal in circumstellar disks formed during the protostellar phase, 
and as such they influence the disk evolution and the subsequent planetary 
system formation processes (Alexander et al. 2014). 
These processes occur on spatial scales (fraction of AU) that
cannot be resolved by present instrumentation. However, selective tracers of the
different phenomena can be identified in the optical/IR spectra of CTT stars and used to 
infer physical and kinematical properties of the relevant regions, provided that sufficient
spectral resolution is adopted.

This is efficiently done with echelle spectrographs that simultaneously cover a wide wavelength range so that
the various phenomena characterizing the active stellar
and circumstellar environment can be simultaneously probed.  This is particularly important for stars
known to be variable, as one can derive a snapshot of all the relevant physical parameters
in a specific phase of the system activity. 
The potential of wide-band optical/infrared spectroscopy has recently been demonstrated 
by several spectral surveys on populations of young stellar objects (YSOs) 
(e.g., Antoniucci et al. 2014; Alcal\'a et al. 2017;
Manara et al. 2017), where this technique has been used to derive a complete characterization 
of the stellar and accretion properties of the stars. 

Wide-band, near-IR (NIR) high-resolution spectroscopy is a much less
explored observational tool for the characterization of YSOs,
despite the presence of several important diagnostic features providing 
complementary information to those in the optical range. 

The power of NIR high-resolution spectroscopy relies in particular
on its unique ability to simultaneously probe the properties of 
gas and dust at different spatial scales in the star-disk interaction regions; the accretion columns and hot spots, the inner gaseous disk, the stellar
and disk winds, and the collimated jets.

For example, the infrared emission in excess to the photospheric spectrum
is directly connected with the properties of the inner disk, and can be inferred 
from a measure of the veiling on the numerous IR absorption photospheric lines, provided that the spectral resolution is high enough
(R $\gtrsim 30\;000$; e.g. McClure et al. 2013, Johns-Krull \& Valenti 2001, Muzerolle et al. 2003,
Fohla \& Emerson 1999). This NIR excess is attributed to the emission of the 
inner edge of the dusty disk,
where the dust is heated by radiation from both the stellar photosphere and accretion shock.
As such, it can provide interesting complementary information with respect to the 
ultraviolet and optical excess, which originate directly from the accretion shock at the stellar
surface (e.g. Calvet \& Gullbring 1998; McClure et al. 2013). 
 
Infrared permitted lines of \hi from the Brackett and Paschen series 
are less affected than Balmer lines by opacity effects and thus can
be more efficiently used to retrieve information on the different components contributing
to their profile (e.g. accretion columns and winds; Folha \& Emerson 2001, Nisini et al. 2004,
Antoniucci et al. 2017, Giannini et al. 2017). 
In addition, the lower extinction with respect to optical lines 
makes the IR \hi\ lines excellent proxies for accretion luminosity
in moderately embedded sources.

Among other permitted lines, the \hei\ 1.08\um\, line is particularly interesting:
It has a high metastable lower level, setting up a favourable situation for tracing 
outflowing gas in absorption. Indeed, the \hei\ 1.08$\mu$m line profiles in CTTs 
often display prominent P-Cygni absorption indicative of ionised winds 
close to the central star (Edwards et al. 2003) . Models show that the shape of this absorption changes
depending on the nature of the wind (Kurosawa et al. 2011, Kwan et al. 2007); disk-winds produce 
a narrow blue-shifted absorption while in stellar winds the absorption covers a 
much larger velocity range. The \hei\ 1.08\um\, line profile thus provides direct pieces of information on
the properties of the inner winds that are not retrievable from optical observations.
The \hei\ 1.08\um\, line is also sensitive to magnetospheric
accretion flows, as it shows redshifted absorptions below the continuum more often that 
other permitted lines (e.g. Edwards et al. 2006).

Finally, IR spectroscopy is able to probe the molecular gas in the inner gaseous
disk and winds,  through H$_2$ ro-vibrational lines and the CO overtone
emission. A spectral resolution of a few \kms\, is needed in order to distinguish their origin 
in gas bound to the disk or from shocked emission in outflows.
 
We report here the first IR spectroscopic observations covering (almost 
completely) the spectral range from 1 to 2.4\um\, at high spectral resolution 
(R$\sim$ 50\,000) of two accretion-active T Tauri stars, namely XZ Tau and DR Tau.
The aim of this work is to characterise the star-disk interaction regions of
the two stars through the analysis of the above described tracers,
eventually putting in relation
the different phenomena probed by the observed spectral features.

The paper is structured as follows: The two sources are described in Sect.~\ref{sec:sources}.
Observations, data reduction, and a discussion on the photometric level of the sources at the moment of the observations are given in Sects.~\ref{sec:observations} and \ref{sec:phot}. Results are presented in Sect.      \ref{sec:results} and the derived veiling and rotational velocities are reported in Sect.~\ref{sec:veiling_rotvel}. Parameters derived from line analysis are presented and discussed in Sects.~\ref{sec:lines}, 8, and 9 and conclusions are finally summarised in Sect.~\ref{sec:conclusions}

\section{Description of the sources}
\label{sec:sources}
XZ Tau and DR Tau are two well studied CTTs that present episodic photometric and spectroscopic variability.
As such they have been included among the variable stars of EXor type (e.g. Lorenzetti et al. 2009).
These are young sources showing erratic increases of brightness with typical time-scales of months/years,
which can be explained in terms of recurrent events of enhanced disk accretion (e.g. Audard et al. 2014, Lorenzetti et al. 2012). 

XZ Tau has displayed quite frequent variations in the recent past, with recurrent increases of the optical brightness up to 2 mag.
The source consists of two components: XZ Tau N (or XZ Tau B) and XZ Tau S (or XZ Tau A), 
separated by 0\farcs 3. The two sources have similar masses (0.37 and 0.29 \msun, respectively) 
but XZ Tau N is the one that shows the richest spectrum in optical emission lines (Hartigan et al. 2003). 
Carrasco-Gonz{\'a}lez et al. (2009) suggested the presence of a third component at about 0\farcs 09
from  XZ Tau S, which was, however, not confirmed by recent ALMA observations (ALMA partnership 2015; Zapata et al. 2015). 
HST observations have shown a series of expanding bubbles around the system: This outflow
is driven by XZ Tau S, as indicated by their alignment with this source (Krist et al. 2008).
ALMA observations have also revealed the presence of a compact and collimated bipolar molecular outflow that is at the base of the optical 
expanding flow (Zapata et al. 2015).
The source shows significant optical-line-profile variability on time-scales of months (Chou et al. 2013).

DR Tau showed a slow but constant brightness increase between 1960 and 1980
of more than three visual magnitudes. After this period, its average brightness has remained high, while showing smaller
variations on time-scales of days. Similarly to XZ Tau, the optical spectrum of DR Tau is
characterised by strong and variable (both in strength and in shape) \hi lines with signatures of both infall and winds (Alencar et al. 2001). 
A strong P-Cygni signature in the \hei\ 1.08\um\, line (Edwards et al. 2006) also testifies for a hot stellar wind.
DR Tau is among the CTTs that present the largest values of the veiling, both in the optical and in the IR,
indicative of  a large accretion luminosity. 
High- velocity blue-shifted [OI]6300\AA\, emission, indicative of a jet, was present in the
spectrum published by Hartigan et al. (1995), but seems to have recently disappeared (Simon et al. 2016).

\section{Observations and data reduction}
\label{sec:observations}

XZ~Tau and DR~Tau were observed in September 2014, 8th and 13th, respectively, with the GIANO spectrograph 
(Oliva et al. 2012, Origlia et al. 2014, Tozzi et al. 2014) placed at the Nasmith A 
focus of the Telescopio Nazionale Galileo (TNG; La Palma, Spain). 
GIANO is a cross-dispersed infrared (0.95-2.4 $\mu$m) spectrograph with a resolving power of $R=50\,000$.  
The data were collected during the Science Verification with the instrument in a provisional
configuration with the cryogenic spectrograph positioned on the rotating floor and far
from the telescope focal plane. 
With this set-up GIANO received the light from the telescope through two fibres (A and B) with a diameter on sky of 1\arcsec and placed at a fixed
projected distance of 3 arcsec. 
The spectra were acquired with the nodding-on-fibre technique, that is, by alternately observing 
the target through fibre A and B (while the other fibre observes the sky) with the same exposure time.
The subtraction of the two exposures ensures an optimal removal of both the sky emission and instrumental background. 
The total exposure time was 30 min for both DR Tau and XZ Tau.

The extraction and wavelength calibration of the spectra were performed following the 2D GIANO data reduction prescriptions\footnote{More details can be found on the GIANO website http://www.tng.iac.es/instruments/giano and in the document available at http://www.bo.astro.it/giano/documents/ new\_GIANO\_cookbook\_for\_data\_reduction.pdf}, as described also in Carleo et al. (2016) and Caffau et al. (2016).

Exposures with a halogen lamp were employed to map the order geometry and for flat-field, while
wavelength calibration was based on lines from a uranium-neon lamp.
The calibration produces an accuracy on the velocity of the order of 0.5 \kms\, as
estimated from the radial velocity analysis discussed in section 4.1.

The final GIANO spectral coverage is complete up to 1.7$\mu$m, whereas it is about 75\% in the $K$-band, because at longer wavelengths the orders become larger than the 2k$\times$2k detector. In particular, the region at 2.16\um\, around the Br$\gamma$ line is not covered.

The one-dimensional spectra resulting from the GIANO pipeline are not corrected for the contribution of telluric features. Therefore, spectra of 
telluric standards were obtained to clean the target spectra from this contribution.
In particular, the spectrum of the star Hip 029216 was employed for XZ Tau and that of Hip 089584 for DR Tau.
After removing the intrinsic features of the standard spectra, we used the  
IRAF\footnote{IRAF is distributed by the National Optical Astronomy Observatory, which is operated by the Association of the Universities for Research in Astronomy, inc. (AURA) under cooperative agreement with the National Science Foundation.} task {\sc telluric}
to correct for the telluric absorptions. The procedure consists of dividing the target spectrum by the telluric spectrum multiplied by an appropriate scaling factor that depends on 
the ratio of the airmass of the target and the telluric standard. This factor is typically close to unity, 
as targets and their assigned 
telluric standards were observed at very similar airmasses.

\section{Photometric state of the two sources}
\label{sec:phot}

\begin{figure}[t]
\includegraphics[angle=0, width =9cm, trim = 0.5cm 0.5cm 0.8cm 0.4cm,clip]{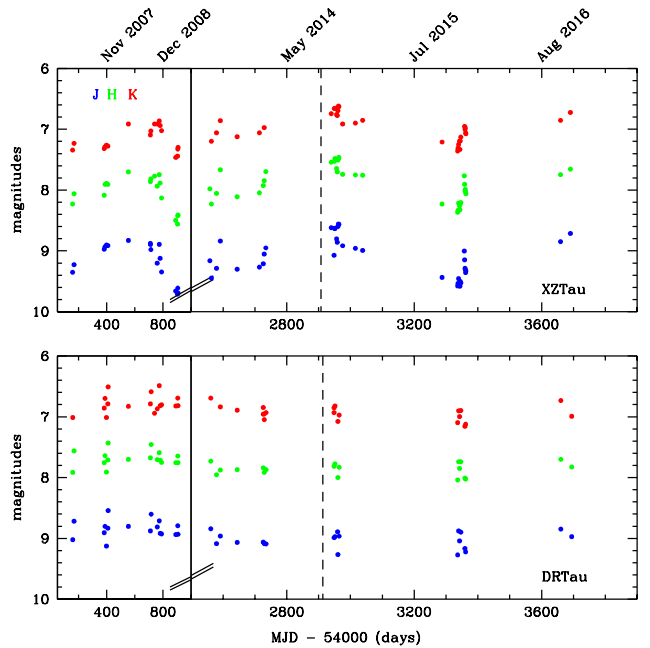}
\caption{\label{fig:phot} Near-infrared ($JHK$ bands) lightcurves of XZ Tau (top) and DR Tau (bottom) based on observations taken within the EXORCISM programme around the date of GIANO observations, which is indicated by the vertical dashed line.}     
\end{figure}

Near-infrared photometry of the targets is not available for the dates of the GIANO spectroscopy. 
However, a few NIR photometric measurements of the two sources, taken with a sparse sampling, were acquired within the EXORCISM programme (Antoniucci et al. 2013; Antoniucci et al. 2014) 
using the SWIRCAM camera mounted on the Campo Imperatore 1.1m AZT-24 telescope (D'Alessio et al. 2000).
The corresponding light curves of XZ Tau and DR Tau during the last few years are shown in Fig.~\ref{fig:phot}.

The mean brightness of XZ Tau was steadily increasing in the period of the GIANO observations. 
We have considered a linear fit over 
the two photometric points before and the two after our observations, thus obtaining a linear averaged trend for that period. The fit gives, at the date of the GIANO observation, $J$=8.69, $H$=7.57, and $K$=6.75, which we assume as the magnitudes of XZ Tau. These values can be compared with the
median near-IR magnitudes (basically indicating the quiescence values) $<J>$=9.12, $<H>$=7.92, and $<K>$=7.05,
obtained from all the available data. 
The registered magnitude difference supports the evidence that the object was in a moderately active accretion 
state at the moment of our observations.

For DR Tau, the closest photometric measurements were acquired 35 days after the GIANO spectroscopy, providing $J$=8.99, $H$=7.81, and $K$=6.85, which are very similar to the median values obtained from the EXORCISM database ($<J>$=8.93, $<H>$=7.79, and $<K>$=6.86). Considering this and that the light curve in Fig.~\ref{fig:phot} shows no clear trend in the period of interest, we may assume that the object was at the same (substantially constant) luminosity level typical of its recent past.

\section{Results}
\label{sec:results}

\subsection{Radial velocity}
\label{sec:rad_vel}

We determined the radial velocities (RVs) of XZ~Tau and DR~Tau, observing with GIANO two 
slowly-rotating non-accreting templates with 
the same spectral types as both targets (i.e. M2 and K7, respectively) and known RVs, namely HD119850 and HD28343. 
The heliocentric RVs of XZ~Tau and DR~Tau were determined through the task 
{\sc fxcor} of the IRAF package {\sc rv}, which cross-correlates the target and template 
spectra, excluding regions affected 
by broad lines or prominent telluric features. The centroids of the cross-correlation function 
(CCF) peaks were determined by adopting 
Gaussian fits, and the RV errors were computed using a procedure that considers 
the fitted peak height and the antisymmetric noise 
(see \citealt{tonrydavis1979}). This procedure was applied for all spectral orders we analyzed for the veiling and rotational 
velocity determinations (Section\,\ref{sec:veiling_rotvel}). 
The weighted averages of the RV measurements obtained for each useful spectral
order are $V_{\rm rad}=18.5 \pm 0.6$ and $22.8 \pm 1.1$ km s$^{-1}$ for
XZ\,Tau and DR\,Tau, respectively; in good agreement with the values obtained by
Nguyen et al. (2012).

\begin{table*}
\label{tab:lines}
\caption[]{Emission line parameters.}
\begin{center}
\begin{tabular}{lcc|ccc|cccc}
\hline
\hline
  &  &     &      \multicolumn{3}{c}{XZ Tau} & \multicolumn{3}{|c}{DR Tau} \\
 $\lambda _{teo}$ & ID & E$_{up}$ & V$_p$ & FWHM & EW & V$_p$ & FWHM & |EW| \\
  (nm)  &  & (eV) &  (\kms)  & (\kms)  & (\AA) & (\kms)  & (\kms)  & (\AA) \\
  \hline\\[-9pt] 
\multicolumn{9}{c}{Hydrogen and Helium lines} \\
\hline
 1005.21   & Pa$\delta$ & 13.32 &  $-$19.5  &178.    & 7.4   & $-$12.8  & 186.  & 6.3  \\
 1094.11   & Pa$\gamma$ & 13.22 & $-$17.6   & 182.   & 8.4   & $-$13.4  & 196.  & 16.3 \\
 1282.16   & Pa$\beta$  & 13.05 &  $-$18.1  & 190.   & 9.9   & $-$13.0  & 242.  & 23.4 \\
 1519.60    & Br 20     & 13.56 & $-$7.4    & 94.    & 4.8   & 13.8   & 110.  & 1.1  \\
 1526.47    & Br 19     & 13.56 &   7.5   & $<$100.    & 0.25  & $-$0.2   & 111.  & 1.2  \\
 1534.60    & Br 18     & 13.56 & $-$29.2   &  $<$150.  & 0.43  & $-$1.2   & $<$140.  & 1.2  \\
 1544.31    & Br 17     & 13.55 &  $-$10.8  & $<$150.   & 0.55  & $-$3.6   & 129.  & 1.4  \\
 1556.07    & Br 16     & 13.54 &  $-$36.1  & 130.   & 0.59  & $-$8.2   & $<$162.  & 1.9  \\
 1570.50    & Br 15     & 13.54 &  15.7   & $<$150.   & 1.1   & $-$2.6   & 130.  & 2.3  \\
 1588.49    & Br 14     & 13.53 &  $-$15.5  & $<$165   & 1.4   & $-$9.8   & 111.  & 2.4  \\
 1611.37    & Br 13     & 13.52 &  $-$10.1  & 121.   & 1.7   & $-$5.7   & 118.  & 2.8  \\  
 1641.17    & Br 12     & 13.50 &  $-$10.8  & 125.   & 1.2   & $-$3.1   & 137.  & 3.1  \\
 1681.11    & Br 11     & 13.49 &  $-$15.0  & 148.   & 2.2   & $-$4.1   & 133.  & 4.1  \\
 1736.68    & Br 10     & 13.46 &  $-$9.5   & 168.   & 3.3   & $-$2.8   & 144.  & 8.8  \\
 1083.20$^a$ & \hei\    & 20.96 &  $-$15.0  & 205.   & 12.9  & 126.3  & 101.  & 4.5  \\
 1278.84    & \hei\     & 24.03 &  $-$2.6   & 154.   & 0.44  & 0.6    & 96.   & 0.3  \\
 2112.58    & \hei\     & 23.58 &  7.6    & 73.3    & 0.42  & 4.7    & 138.  & 0.7  \\
\hline\\[-9pt] 
\multicolumn{9}{c}{Other permitted lines} \\
\hline 
 1000.03  & \ion{Ca}{i}    & 5.98 & $-$10.3 & 149.  & 0.98 & ...& &\\
 1012.66  & \ion{C}{i}     & 9.76 &  4.8 & 134.   & 0.43 & 8.3 &93.    & 0.33\\
 1045.93  & \fei\          & 7.38 &  $-$0.1 & 122.  & 0.56 & ...& &\\
 1046.10  & \fei\          & 6.74 &  7.7  & 57.   & 0.24 & ...& &\\
 1050.40  & \fei\          & 7.45 & $-$1.8 & 136.  & 0.66 & 7.5& 92.    &0.27\\
 1068.60  & \ion{C}{i}     & 8.64 &  $-$15.4 & 131. & 1.3 & $-$3.4& 65. & 0.70\\
 1068.83  & \ion{C}{i}     & 7.57 &  $-$9.8 &  63. & 0.14 & $-$3.4 & 38. & 0.20\\
 1069.42  & \ion{C}{i}     & 8.65 &  5.5 & 117.   & 1.3 & 3.2& 87.    & 1.5\\
 1071.02  & \ion{C}{i}     & 8.64 &  $-$2.3 & 84.   & 0.50 & 8.7& 72.    & 0.73\\
 1073.25  & \ion{C}{i}     & 8.64 &  $-$15.0 & 104. & 0.81 &0.9 & 73.    & 0.43\\
 1087.34  & \fei\          & 6.76 &  8.2 &  53.   & 0.29 & ...& &\\
 1087.25  & \ion{Si}{i}    & 6.22 &  $-$25.9 & 58.  & 0.43 & ...& &\\
 1091.67  & \fei\          & 7.37 &  $-$6.0 & 89.  & 0.52 &... & &\\
 1128.94  & \ion{O}{i}     & 12.09 &   6.6 & 111. & 4.5 & ...& &\\
 1162.60  & \fei\          & 6.65 &  $-$24.0 & 133.& 0.42 &... & &\\
 1163.20  & \ion{C}{i}     & 9.71 &  $-$13.8 & 136.& 0.82 &... & &\\
 1175.14  & \ion{C}{i}     & 9.69 &  $-$9.0 & 76.  & 0.19 &... & &\\
 1175.57  & \ion{Ca}{i}    & 6.10 &  15.1 & 79.  & 0.50 &22.6 &86.    &0.79\\
 1175.75  & \fei\          & 7.33 &  11.4 & 81.  & 0.20 &... & &\\
 1183.04  & \fei\          & 5.63 &  $-$5.8 & 80.  & 0.46 & ...& &\\
 1183.31  & \fei\          & 6.67 &  $-$2.9 & 38.  & 0.14 & ...& &\\
 1184.19  & \fei\          & 6.66 &  0.9 & 103.  & 0.62 &... & &\\
 1189.93  & \fei\          & 6.68 &   $-$17.6 & 57.& 0.24 & ...& &\\
 1195.53  & \fei\          & 7.40 &  $-$21.1 & 37. & 0.07&... & &\\
 1197.38  & \fei\          & 6.05 &  12.4 & 104. & 0.53  & ... &  &\\
 1197.63$^b$  & \fei\      & 3.21 & 44.9 &33.& 0.12 & ...& &\\
 1198.75  & \ion{Si}{i}    & 5.96 &  $-$20.3 & 60.  & 0.31 &... & &\\
 1198.86  & \fei\          & 5.61 &  10.7 & 32.   &  0.12 &... & &\\ 
 1199.37  & \fei\          & 6.43 &  12.8 & 48.    & 0.09 &... & &\\
 1199.48$^b$ & \ion{Si}{i} & 5.95 &  38.1&37. & 0.11 &... & &\\
 1203.48  & \ion{Si}{i}    & 5.98 &  $-$10.7 & 99. & 0.52  &... & &\\
 1203.50  & \fei\          & 6.08 & 0.2  & 62.   & 0.30 &... & &\\
 1208.53  & \ion{Si}{i}    & 7.29 &  8.8 & 36.    & 0.10  &... & &\\
 1208.85  & \ion{Mn}{i}    & 6.88 &  $-$0.8 & 33.  & 0.09 &... & &\\
 1261.76  & \ion{C}{i}     & 9.83 & 9.3 & 72.     & 0.13 &$-$31.8 &90.  &0.1\\
\hline
\end{tabular}
\end{center}
\tablefoot{
\tablefoottext{a}{Line parameters refer only to the emission component.}
\tablefoottext{b}{Identification uncertain as the radial velocity is too different from the average of the other lines.}\\
}
\end{table*}

\subsection{Observed emission features}

The XZ Tau and DR Tau GIANO spectra are dominated by strong emission lines 
of \hi\ (Brackett and Paschen), \hei, and other metallic neutral lines.
Identification of the permitted lines were performed using the Atomic line list\footnote{http://www.pa.uky.edu/~peter/atomic} and the NIST\footnote{http://physics.nist.gov/PhysRefData/ASD/lines\_form.html} databases. 
Table 1 lists the major features, including the identifications, 
equivalent widths (EWs), and kinematical parameters (i.e. peak velocity $V_p$ and FWHM ) 
derived by a Gaussian fitting of the emission line. Velocities were corrected for the 
stellar radial velocity discussed in Section~\ref{sec:rad_vel}. 

The upper panel of Table 1 includes
the \hi\ and \hei\ lines, while the rest of the table lists all the other permitted lines.
Forbidden lines, for example \feii\ at 1.25 and 1.64\um , remain below the detection limit 
in both objects, while few H$_2$ v=1-0 ro-vibrational transitions are observed 
in XZ Tau (Table~\ref{tab:lines}). 

As for the XZ Tau system, Hartigan et al. (2003), who observed the spectra of the XZ Tau A and B
components separately, found that XZ Tau B is the richest in bright permitted emission lines. We therefore
discuss our spectrum as part of the hypothesis that it is mainly dominated by the XZ Tau B
contribution.

\subsection{Accretion luminosity and mass-accretion rates}

We used the estimated $J$-band magnitude of the two stars (see Sect.~\ref{sec:phot})
to flux-calibrate the Pa$\beta$ spectrum and use it to derive the accretion luminosity
of the two sources. 
The derived Pa$\beta$ fluxes are 1.1$\times$10$^{-12}$ and 1.9$\times$10$^{-12}$ erg\,s$^{-1}$\,cm$^{-2}$ for XZ Tau 
and DR Tau, respectively.
We then converted the flux into a line luminosity adopting for XZ Tau \av\ = 1.4 mag (Hartigan \& Kenyon 2003) and 
for DR Tau \av\ = 1.85 mag (Banzatti et al. 2014), and used the line-accretion luminosity relationship given in Alcal\'a et al. (2014) to obtain $L_{acc}$ = 0.34 
and 0.79 \lsun, respectively.

For XZ Tau, assuming that most of the accretion luminosity is derived
from XZ Tau B with a mass of 0.4 \msun\, and a radius of 1.2 R$_\sun$ 
(Hartigan \& Kenyon 2003), we estimate a mass accretion
rate \macc = 4.2$\times$10$^{-8}$ \msunyr.
For comparison, we note that the Pa$\beta$ flux that we obtain in XZ Tau and that we used to 
compute \macc\ is about a factor two higher than the flux measured by 
Lorenzetti et al. (2009) in 2007, during a high state of the object.

For DR Tau, we adopt $M_\star$ = 0.75 M$_\sun$ and 
$R_\star$ = 2.2 R$_\sun$ (e.g. Rigliaco et al. 2013), deriving 
\macc = 8.4$\times$10$^{-8}$ \msunyr.
Banzatti et al. (2014) monitored the DR Tau accretion luminosity variations during 
November 2012-January 2013, finding $L_{acc}$ variations between 1.5 
and 3.5 \lsun , corresponding to variations
of \macc\, between 1.6 and 3.9$\times$10$^{-7}$\msunyr. Lorenzetti et al. (2009),
in their monitoring during a one year period 
between 2007 and 2008, measured $L_{acc}$ variations between 0.9 and 2.4 \lsun .
It seems therefore that \lacc\, variations up to a factor of five on timescales
of months are common in DR Tau, and that we have now monitored a period
of relatively low accretion activity of this star.

In conclusion, both sources are CTT stars that display high accretion luminosity values, with  
XZ Tau showing a \lacc\ higher than its mean level, and DR Tau showing a lower than normal \lacc, albeit still greater than the one of XZ Tau.

\begin{table}[t]  
\caption{Veiling values for both targets at several wavelength ranges.}
\label{tab:veiling}
\begin{center}  
\begin{tabular}{c|cc|cc}
\hline
\hline
\noalign{\smallskip}
  Orders  & \multicolumn{2}{c}{XZ Tau} & \multicolumn{2}{|c}{DR Tau} \\
  $\lambda$ range & $r_\lambda $ & $v\sin i$ &  $r_\lambda$ &  $v\sin i$  \\
   ($\mu$m) & & (\kms) & &  (\kms)   \\ 
\noalign{\smallskip}
\hline
\noalign{\smallskip}
0.958-0.980 & 0.6$\pm$0.3 & 15.0&                 ...    &   ...       \\ 
0.971-0.992 & 0.2$\pm$0.3 & 21.5&                 ...    &   ...       \\ 
1.165-1.191 & 0.7$\pm$0.2 & 16.5&  3.3    $\pm$   0.5    &     $<$9    \\ 
1.242-1.270 & 0.4$\pm$0.3 & 18.5&  3.1    $\pm$   0.6    &     $<$7    \\ 
1.284-1.313 & 1.0$\pm$0.2 & 20.0&  4.3    $\pm$   0.6    &     $<$12   \\ 
1.547-1.581 & 1.2$\pm$0.3 & 18.0&  5.1    $\pm$   0.6    &      6.5    \\ 
1.579-1.614 & 0.5$\pm$0.3 & 20.0&  4.6    $\pm$   0.2    &      6.5    \\ 
1.648-1.684 & 1.0$\pm$0.2 & 18.0&  4.2    $\pm$   0.5    &     $<$10   \\ 
1.723-1.761 & 1.3$\pm$0.3 & 16.5&                 ...    &   ...       \\ 
2.106-2.153 & 2.8$\pm$0.3 & 13.0&  8.5    $\pm$   0.2    &     $<$13   \\ 
2.230-2.280 & 3.0$\pm$0.2 & 19.5&  >6.5                  &   ...       \\ 
2.298-2.349 & 2.6$\pm$0.3 & 14.5&  >8                    &   ...       \\
\noalign{\smallskip}
\hline
\end{tabular}
\end{center}
\end{table}  

\section{Veiling and projected rotational velocity}
\label{sec:veiling_rotvel}

\begin{figure*}
\centering
\includegraphics[width=8cm]{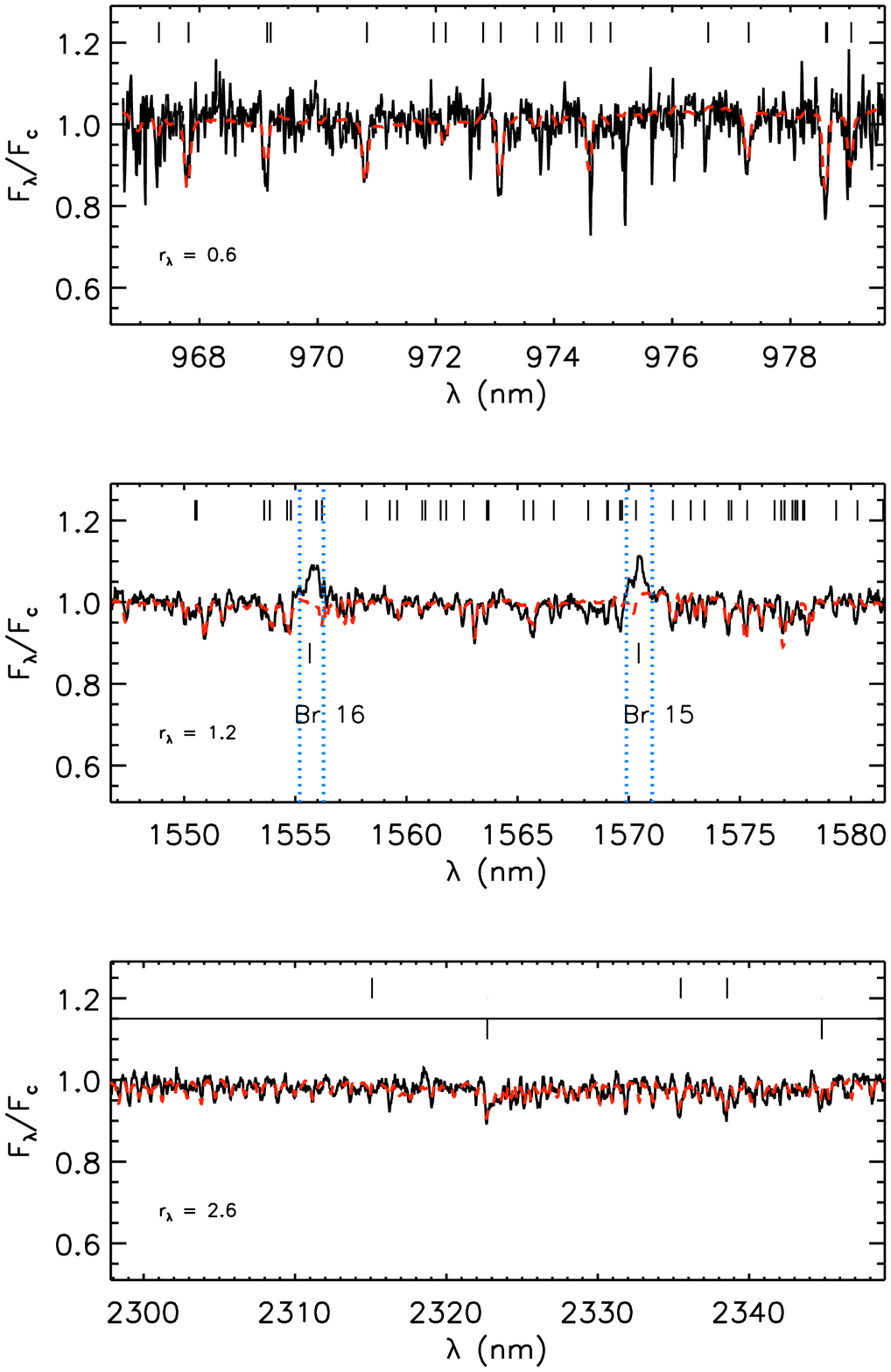}
\includegraphics[width=8cm]{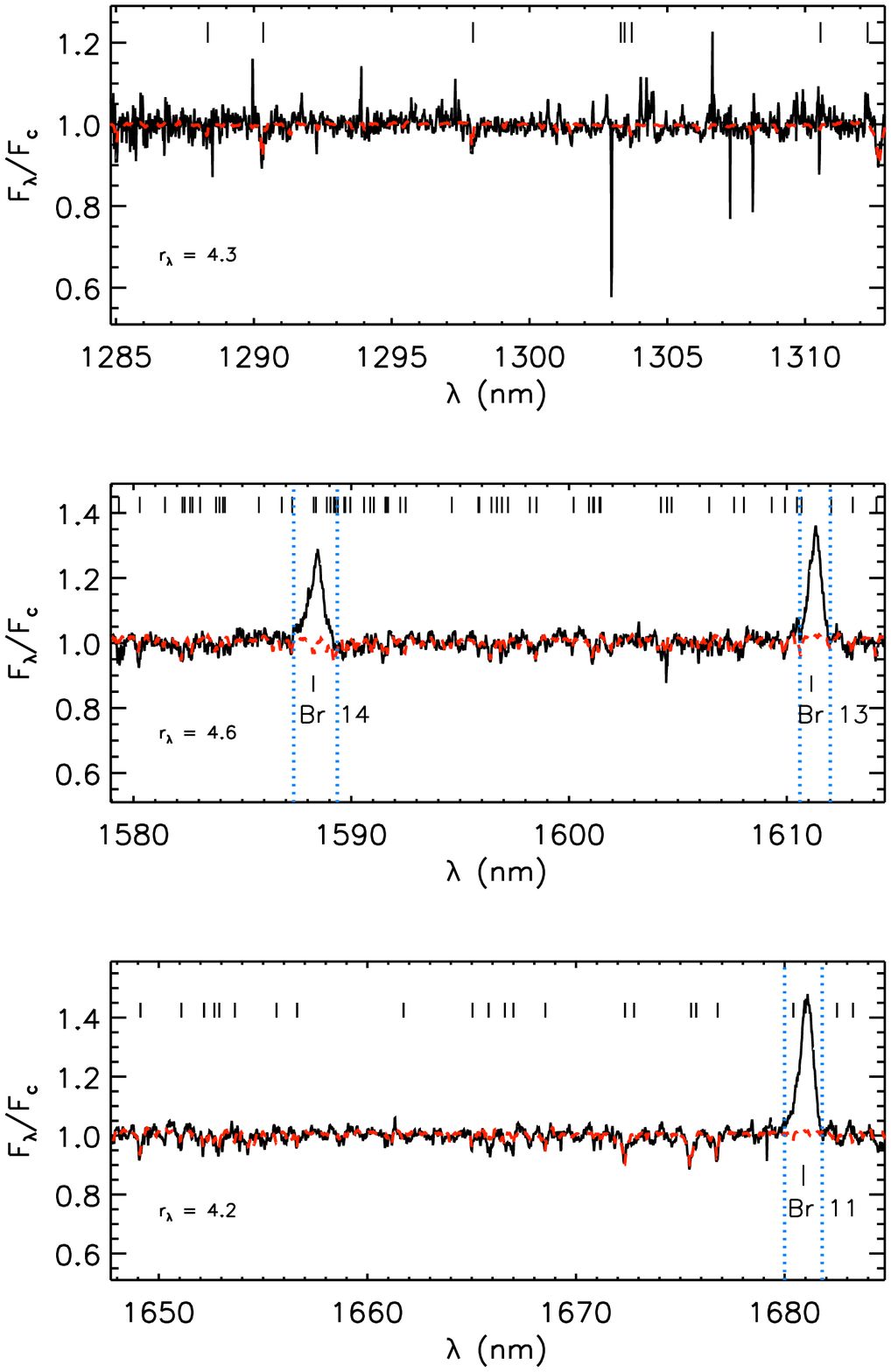}
\caption{Left: Three portions in the $YHK$ bands of the GIANO spectrum of XZ~Tau (solid black line) 
with the rotationally-broadened and veiled spectrum of the M2 template (dashed red line) 
overplotted in the same wavelength ranges.
Absorption features, mainly of the iron-peak group, are 
indicated with short lines. We refer interested readers to \cite{rayneretal2009} and 
\cite{sharonetal2010} for more details about 
absorption features. 
In the bottom panel, $^{12}$CO and $^{13}$CO molecular 
bands are represented by a horizontal line.
Main emission features are marked and excluded from the analysis.
Right: The same for DR Tau in three different wavelength regions, with the 
dashed red line representing the K7 template.}
\label{fig:spectra_veiling}
\end{figure*}

\begin{figure*}
\centering
\includegraphics[width=8cm]{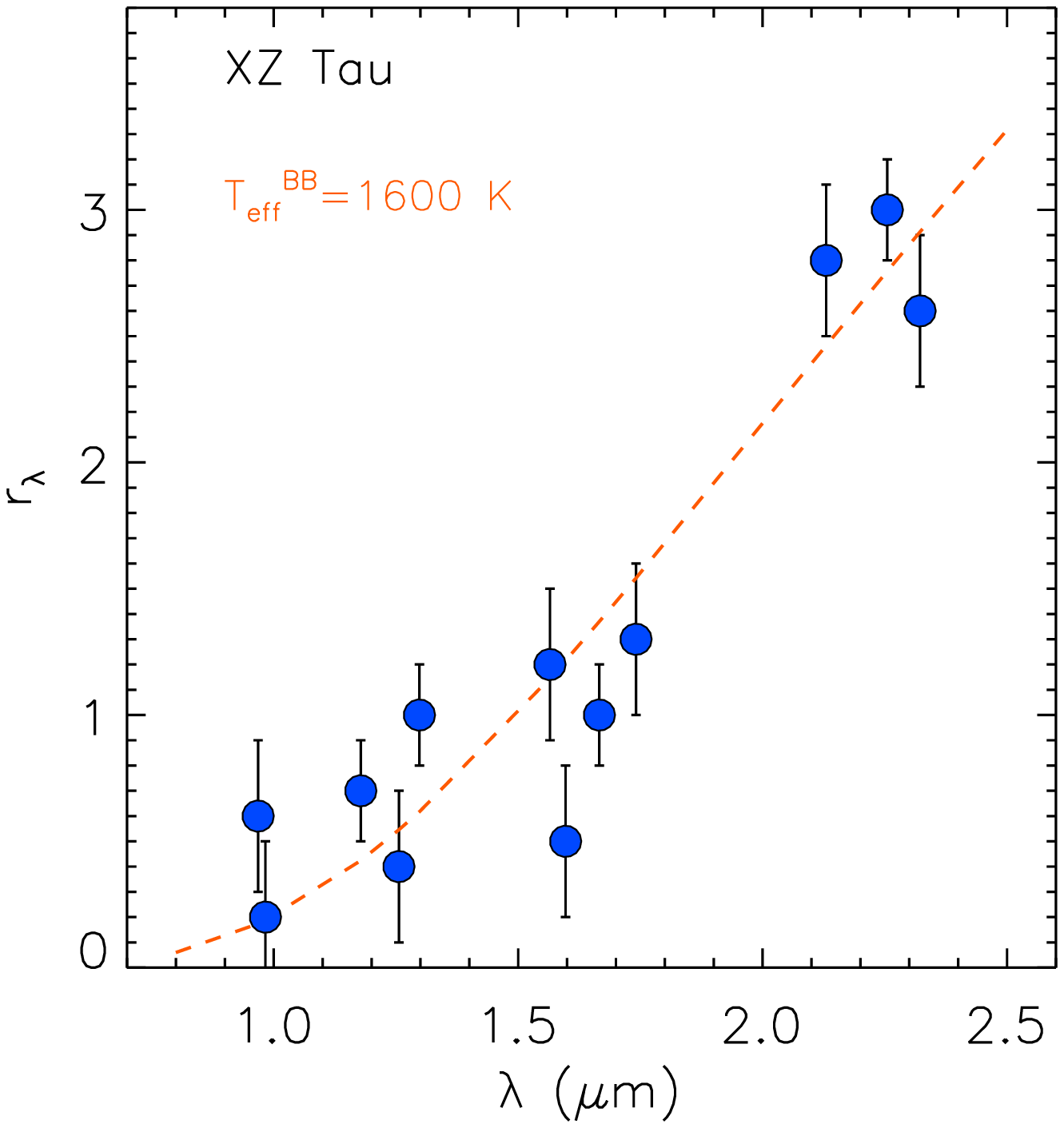}
\includegraphics[width=8cm]{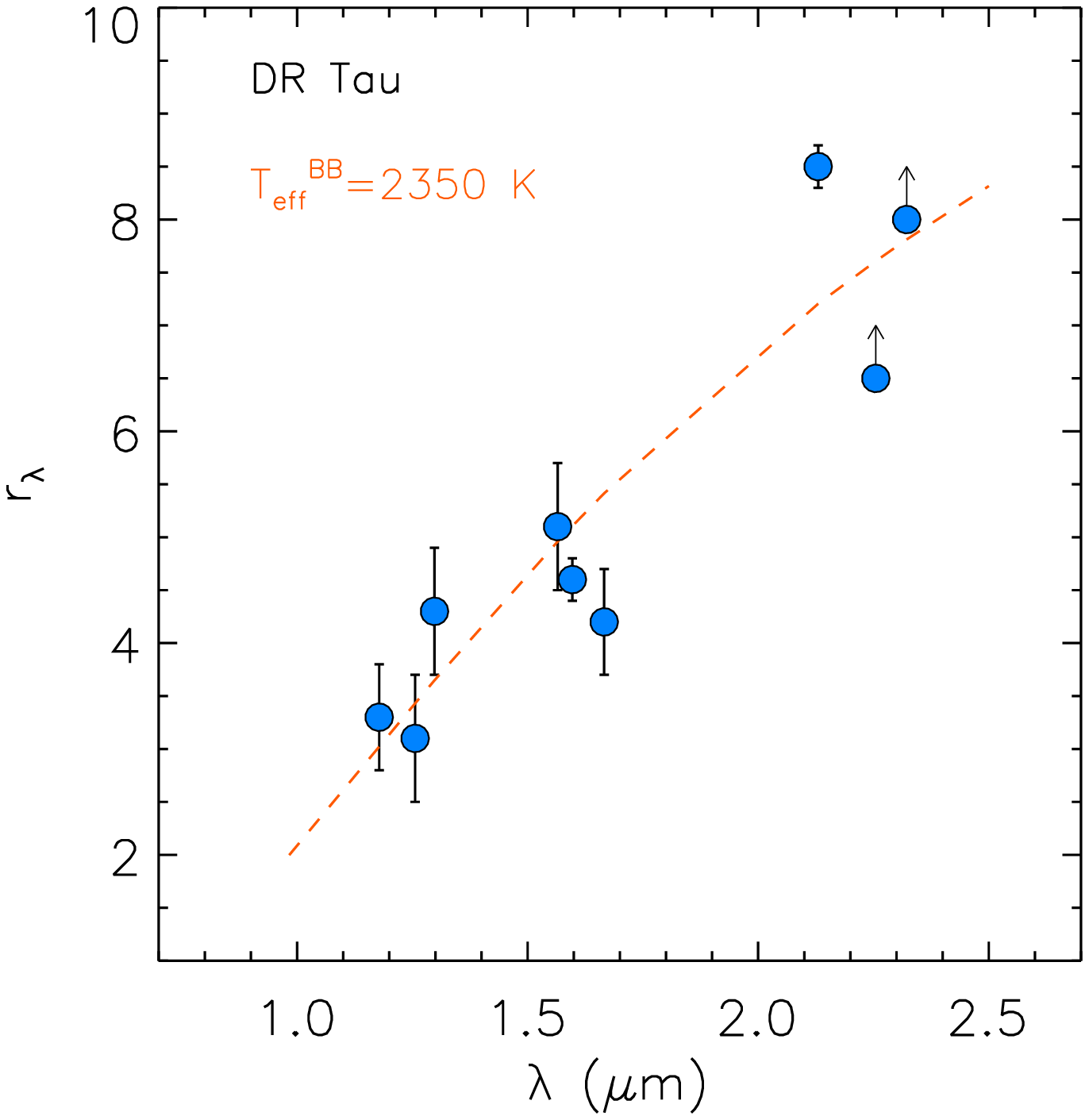}
\caption{Veiling as a function of wavelength for both targets. Error bars refer 
to $\pm$1$\sigma$ of residuals. Red dashed line represents the ratio of blackbodies 
at the photospheric and boundary layer temperatures (see text).}
\label{fig:veiling_lambda}
\end{figure*}

   \begin{figure*}[t!]
      \centering
\includegraphics[angle=0, width =16cm, trim = 0 1cm 0 0, clip]{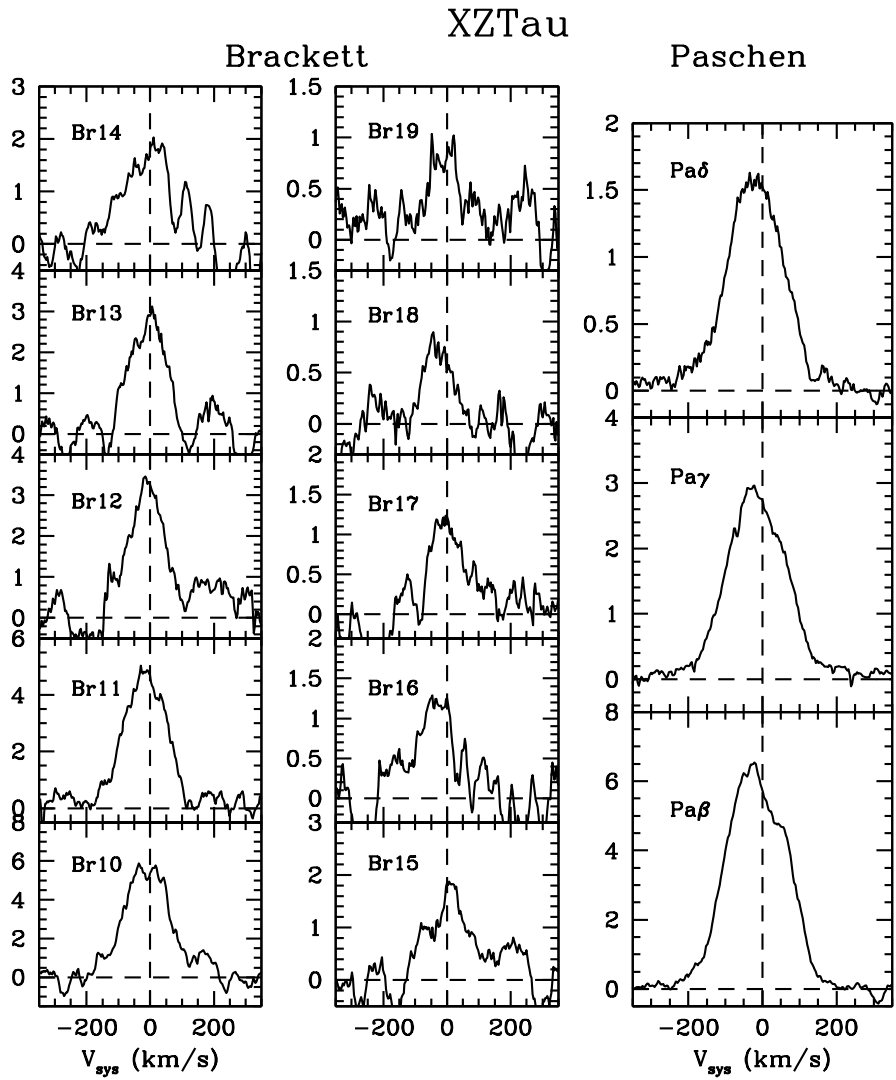}
      \caption{Continuum subtracted spectra of the \hi\ lines detected in XZ Tau. Flux is in arbitrary units.
      }
         \label{HI_XZ}
   \end{figure*}
   
   \begin{figure*}[t!]
      \centering
\includegraphics[angle=0, width =16cm, trim = 0 1cm 0 0, clip]{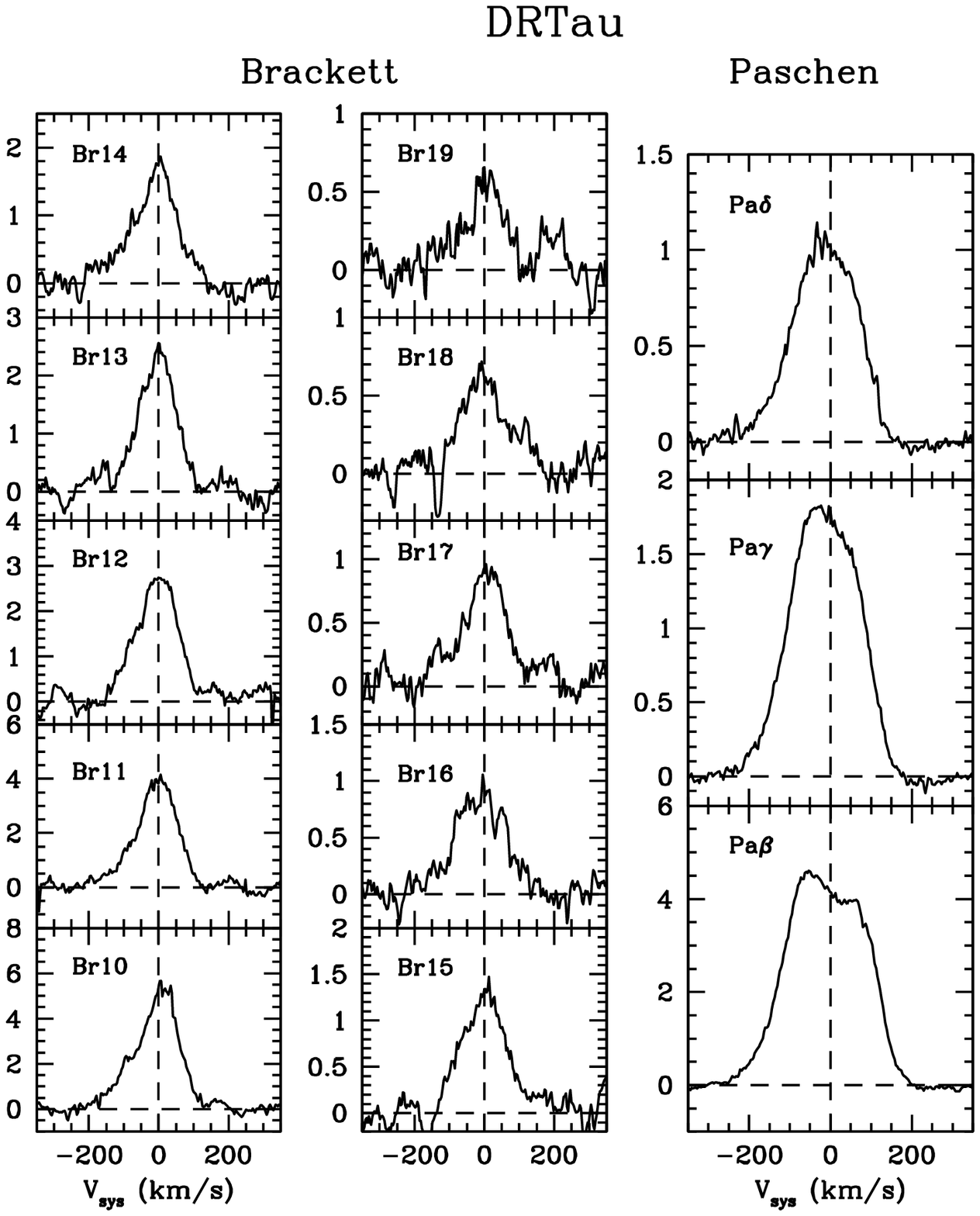}
      \caption{The same as Fig. \ref{HI_XZ} but for DR Tau.
      }
         \label{HI_DR}
   \end{figure*}

In addition to the emission lines, the two spectra show several absorption lines characteristic of
late-type photospheres. These features have been used to derive the projected rotational velocity ($v \sin i$) 
and veiling ($r_\lambda$) of the targets, fixing their spectral types to the 
values recently found in the literature (i.e. M2 and K7 for 
XZ~Tau and DR~Tau, respectively (\citealt{hartigankenyon2003, 2006ApJ...646..319E}).
The veiling can in turn be used to
infer the properties of the emission in excess to the stellar photosphere 
originating in the inner disk.

To this aim, we developed an interactive IDL\footnote{IDL (Interactive Data Language) is a 
registered trademark of Exelis Visual Information Solutions.} 
routine that compares, for each spectral order, the target spectra with template 
spectra of comparable spectral types of slowly rotating, non-accreting stars (i.e. with 
no intrinsic continuum veiling) conveniently 
acquired with the same spectrograph. Before determining the required 
level of continuum excess, the removal of telluric absorption lines was 
performed both for target and templates. The spectra of Hip103087 and Hip029147 were used to remove the telluric contribution from the spectra of HD119850 and HD28343, respectively.
Then, each template was aligned in wavelength with the target spectrum via 
cross-correlation and artificially broadened by convolution with a rotational profile of 
increasing $v \sin i$ (\citealt{Gray2005}) until the minimum residuals were 
obtained. Appropriate linear limb-darkening coefficients by 
\cite{claretbloemen2011} were taken into account within the rotational profile and 
for each spectral region. The code evaluated the degree 
of continuum veiling of the target stars considering the following equation: 
\begin{equation}
\left(\frac{F_\lambda}{F_C}\right) ^{\rm veil} = \frac{\frac{F_\lambda^{\rm phot}}{F_C^{\rm phot}} + r_\lambda}{1 + r_\lambda}\,, 
\end{equation}
\noindent{where $F_\lambda^{\rm phot}$ and $F_C^{\rm phot}$ represent 
the line and continuum fluxes of the underlying photosphere, respectively. 
The template spectra were then artificially veiled by adding an excess 
continuum (leaving the parameter $r_\lambda$ free to vary) 
until the depth of photospheric features matched the depth of the 
features in the target spectra and the minimum residuals were found. 
We assumed $r_\lambda$ constant within each spectral order. A similar procedure was followed also by other authors 
(e.g. \citealt{whitehillenbrand2004, 2006ApJ...646..319E, frascaetal2015}). }
We were able to perform the described procedure for 12 spectral orders for XZ Tau and 9 for DR Tau, owing to the larger veiling of DR Tau.
In Fig.~\ref{fig:spectra_veiling} we show examples of three spectral 
regions in the $YHK$ bands of the XZ~Tau and DR~Tau 
spectra where the veiling was computed. 
The veiling values measured in the considered spectral orders are listed in Table~\ref{tab:veiling} 
for both targets, together with the rotational velocity. 
Our procedure implicitly assumes a solar metallicity  for both the templates and the sources.

The mean $v \sin i$ of XZ~Tau is 17.6$\pm$2.5 km/s, which is in agreement with recent findings (\citealt{nguyenetal2012}). 
Edwards et al. (2006) measured the veiling at 1\um\ in XZ Tau finding it consistent with 0.
We measure $r_{1\mu m} \sim$ 0.4 by averaging the values found in the two ranges
around 1 $\mu m$. This finding indicates that the veiling might have changed in the last years and that 
XZ Tau was probably in a phase of increasing IR excess during our more recent observations. 
Hartigan et al. (2003) observed the optical spectra and measured the 
veiling at 8115\AA\, separately for the two stars of
the XZ Tau system. They found that the primary component (XZ Tau B, that they call XZ Tau p) is the richest
in optical lines and has the largest optical excess. They measured $r_{8115\AA}$ $\sim$ 1.3 for XZ Tau B and
0.06 for XZ Tau A. On this basis, we suggest that the increased veiling value that we measure, with respect
to the Edwards et al. (2006) determination, is dominated by
the excess of XZ Tau B, which seems to be the most active of the two stars. 

For DR Tau, we were able to obtain a $v \sin i$ $\sim6.5$ \kms for two spectral orders
(in agreement with the value of 6.26$\pm$0.12 \kms measured by Nguyen et al. 2012), and upper limits for
the other orders.
The veiling in DR Tau has been measured in the past by several authors who
estimated different values at different epochs. In particular, the veiling
given in the literature ranges between 2 and 3.5 at 1\um\, and between 2.5 and 6 at 
about 1.5\um\, (Edwards et al. 2006, McClure et al. 2013, Fischer et al. 2011).
We find values closer to the high veiling determinations measured by 
Fisher et al. (2011).

For both targets, the resultant excess flux appears to increase with wavelength throughout the observed range 0.96$-$2.4\,$\mu$m 
(see Fig.~\ref{fig:veiling_lambda}). The shape of the $r_\lambda-\lambda$ plot implies emission from optically thick material at a characteristic 
temperature. We were able to qualitatively match the shapes using single-temperature blackbodies at $T \sim 1600$\,K for XZ Tau and  $T \sim 2350$\,K for DR Tau, with emitting areas $\sim29$ and $\sim28$ times greater than the central star, respectively.
The derived blackbody temperatures are in the range found in other T Tauri stars 
(e.g. Muzerolle et al. 2003, McClure et al. 2013). 

The temperature estimated for XZ Tau is consistent with an origin of the emission excess in the inner rim of
the dusty disk, where dust starts to photoevaporate. 

The higher blackbody temperature of DR Tau, however, is incompatible 
with dust emission in the inner disk, as dust is destroyed at a temperature of around 1500 K. 
On the other hand, the derived filling factor exceeds the size of the star, 
excluding that the origin of the excess is the warm post-shocked gas associated with
accretion hot spots. A possible source of emission excess could be 
found in a thick gaseous disk inside the dust sublimation radius, as already         also suggested
by Fischer et al. (2011).
This scenario is consistent with the dust inner radius of $\sim$ 0.1 AU measured
in DR Tau from interferometric observations (Eisner et al. 2005). Indeed, we derive 
an emission radius about 5.3 times larger than the stellar radius, that is, 
$\sim$ 0.05 AU if we adopt $R_\star$ = 2.2 R$_\sun$ (Rigliaco et al. 2013).

\begin{figure}[h]
      \centering
      \includegraphics[angle=0, width =7cm]{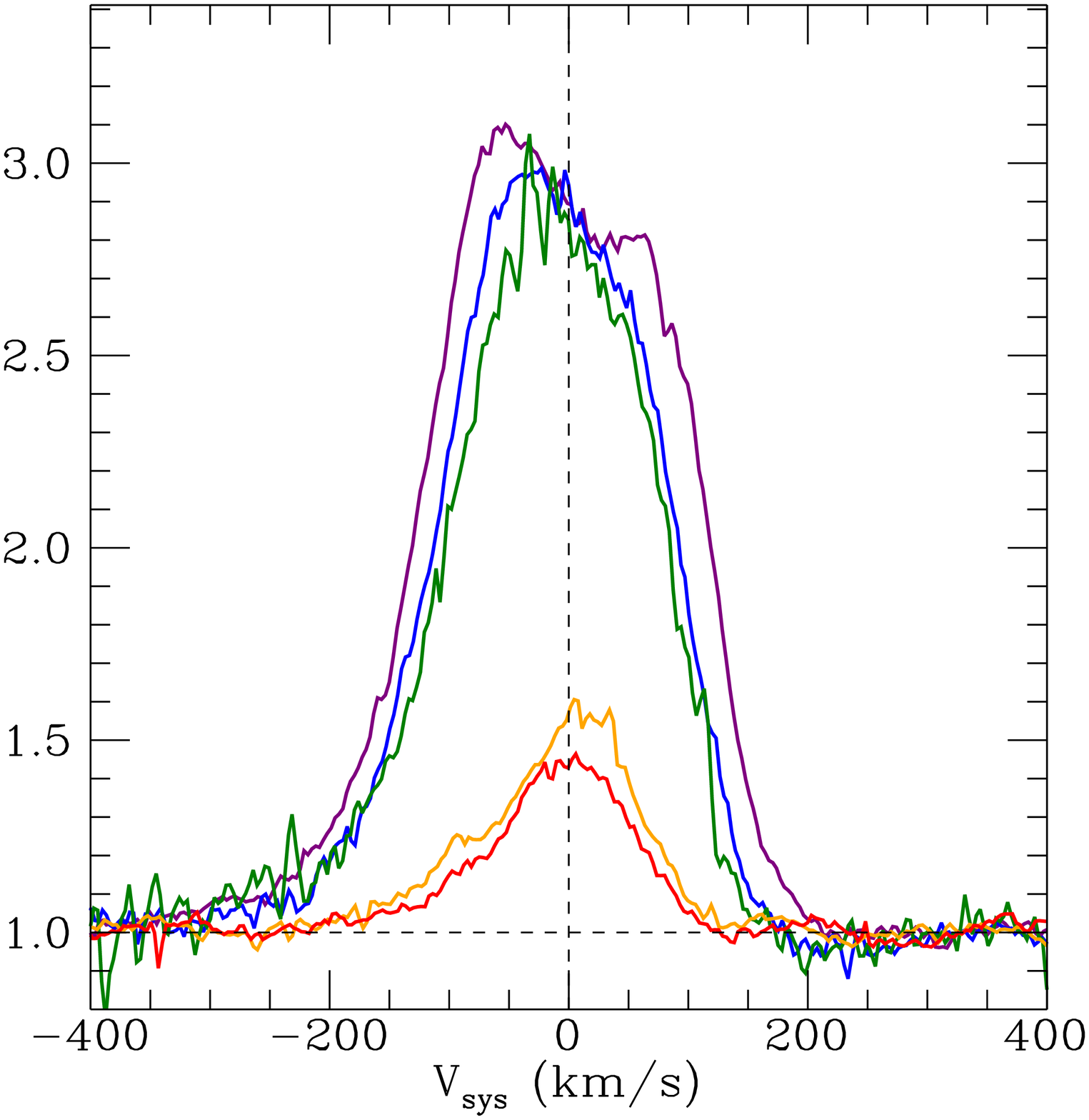}
      \includegraphics[angle=0, width =7cm]{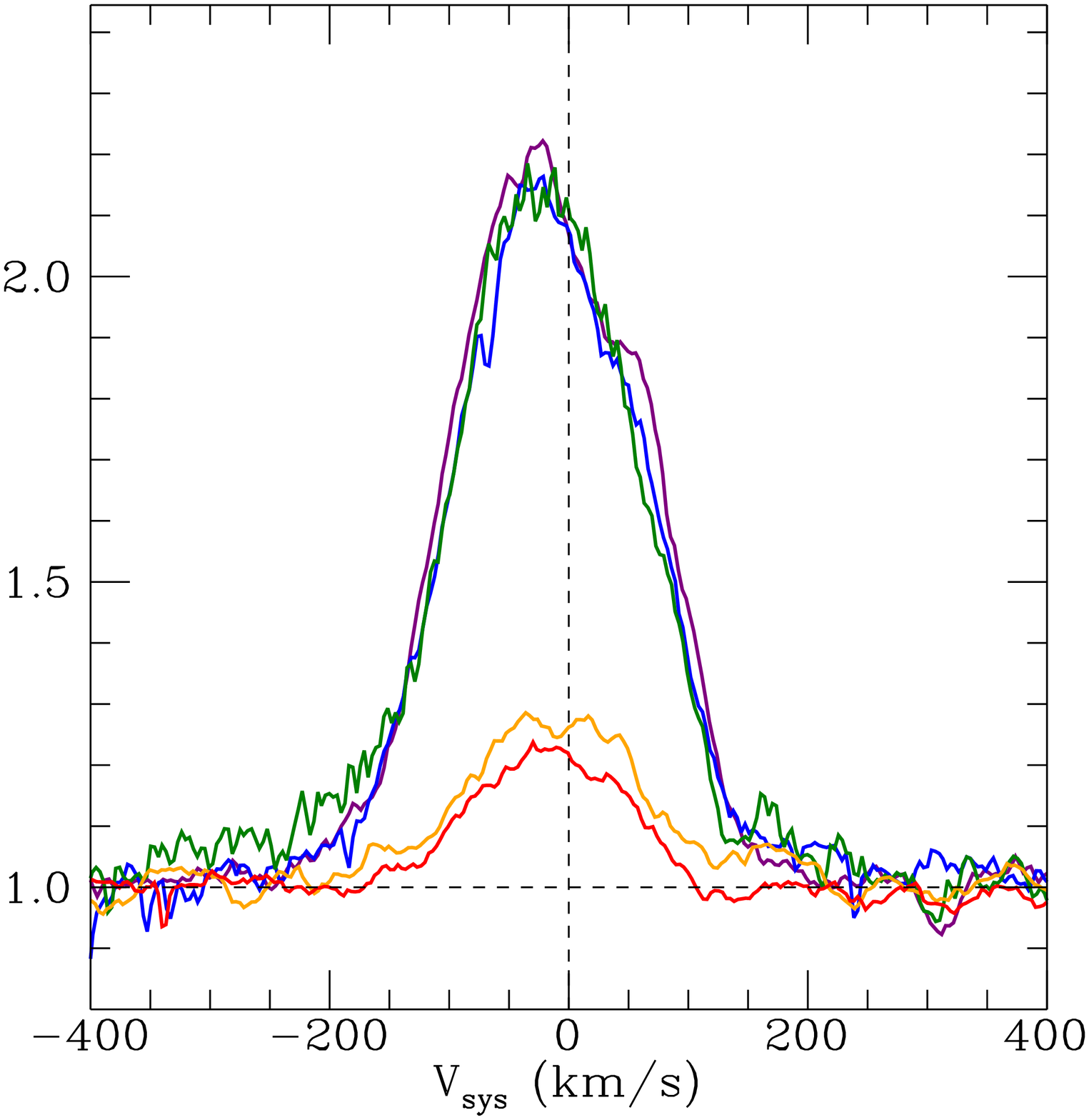}
      \caption{Continuum-normalised \hi\ line profiles in DR Tau (top) and XZ Tau (bottom). The following lines are depicted: Pa$\beta$ (purple), Pa$\gamma$ (blue), Pa$\delta$ (green), 
      Br10 (orange) and Br11 (red).}
      \label{fig:profiles}
\end{figure}

\begin{figure}[h]
      \centering
        \includegraphics[angle=0, width =15cm, trim = 0cm 1.5cm 0cm 3cm, clip]{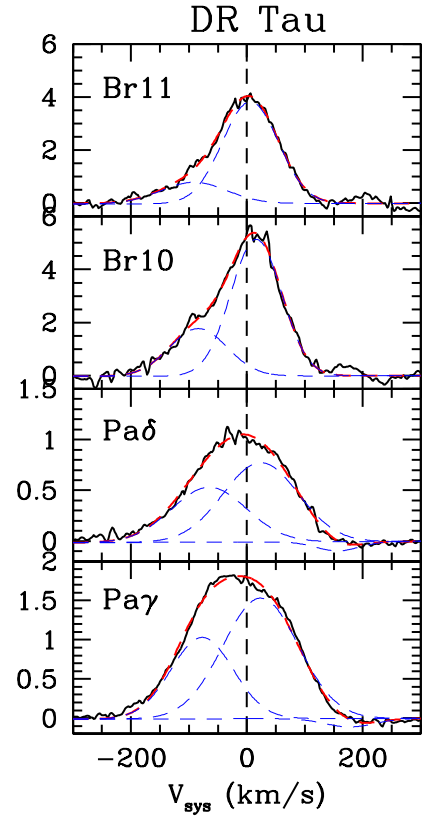}
      \caption{The profile of bright HI lines in DR Tau is deconvolved with two Gaussians, one centred
      close to systemic velocity and the other blue-shifted at V$_{sys}$ $\sim$ $-$70/$-$100\kms .}
      \label{brfit}
\end{figure}

\section{Properties of the permitted lines}
\label{sec:lines}

\subsection{Hydrogen lines}

The strongest emission lines in the observed spectra are the hydrogen Paschen (from Pa$\beta$ to
Pa$\gamma$) and Brackett (from Br10 to Br20) lines. We remind that Br$\gamma$
is not observed as it falls in a hole of the GIANO spectral coverage. 

Hydrogen line profiles are displayed in Figs.~\ref{HI_XZ} and \ref{HI_DR} for XZ Tau and DR Tau, respectively,
while Fig.~\ref{fig:profiles} shows a superposition of bright HI lines to better highlight 
 differences among different line profiles. 
The profiles show some similarities between the two sources. 
In both sources the Paschen lines are the most asymmetric, with asymmetries in the red-shifted part of the line that 
are more prominent in the Pa$\beta$ transition. In DR Tau, an inverse-P Cygni
profile is detected in the Pa$\gamma$ and Pa$\delta$ lines. Such a feature is 
observed also in the H$\delta$ line while it is absent in the
H$\alpha$ and H$\beta$ lines (Petrov et al. 2011, Alencar et al. 2001)
which, on the contrary, are strongly peaked at red-shifted velocities. 
The high-n Brackett lines are more symmetric, consistent with their lower optical depth
with respect to line emission from lower levels (e.g. Antoniucci et. al. 2017, Nisini et al. 2005).
In DR Tau, both Paschen and Brackett lines present a blue-shifted excess 
suggesting the presence of two gas components. Blue-shifted wings are detected
also in the Paschen lines of XZ Tau, but they are less evident in the Brackett lines,
perhaps due to a worse line to continuum ratio in these lines.

In the case of DR Tau, we have deconvolved the lines where this excess is more prominent, 
namely the Pa$\gamma$, Pa$\delta$, Br 10 and Br11, using two main Gaussian emission components and a red-shifted absorption component for
the Paschen lines (see Fig. \ref{brfit}). 
One of the emission components is marginally red-shifted ($\sim$ 10-20 \kms) and has widths of about
100-150 \kms, while the second emission component is blue-shifted ($\sim$ $-$70-90 \kms)
and slightly broader ($\sim$ 110-150 \kms). This blue-shifted component could be related
to the wind at the origin of the absorption in the \ha\, line (Alencar et al. 2011).
This feature shows large variability in both equivalent width (from absorption to 
emission) and velocity peak (from about $-$60 to $-$200 \kms). 
Alencar et al. (2011) suggest that this variability in the blue side of the profile 
can be caused by a sporadic occurrence of high-velocity outflows coupled with geometrical effects. 
We return to the wind variability in Sect.~\ref{sec:hei}.

The observed composite profiles, with red-shifted asymmetries and P-Cygni direct and inverse 
features are qualitatively predicted by models that combine excitation in
accretion columns and disk winds (e.g. Muzerolle et al. 2001, Kurosawa et al. 2011).

\begin{figure*}[t]
      \centering
      \includegraphics[angle=0, width =14.5cm,trim = 0 1.5cm 0 4cm,clip]{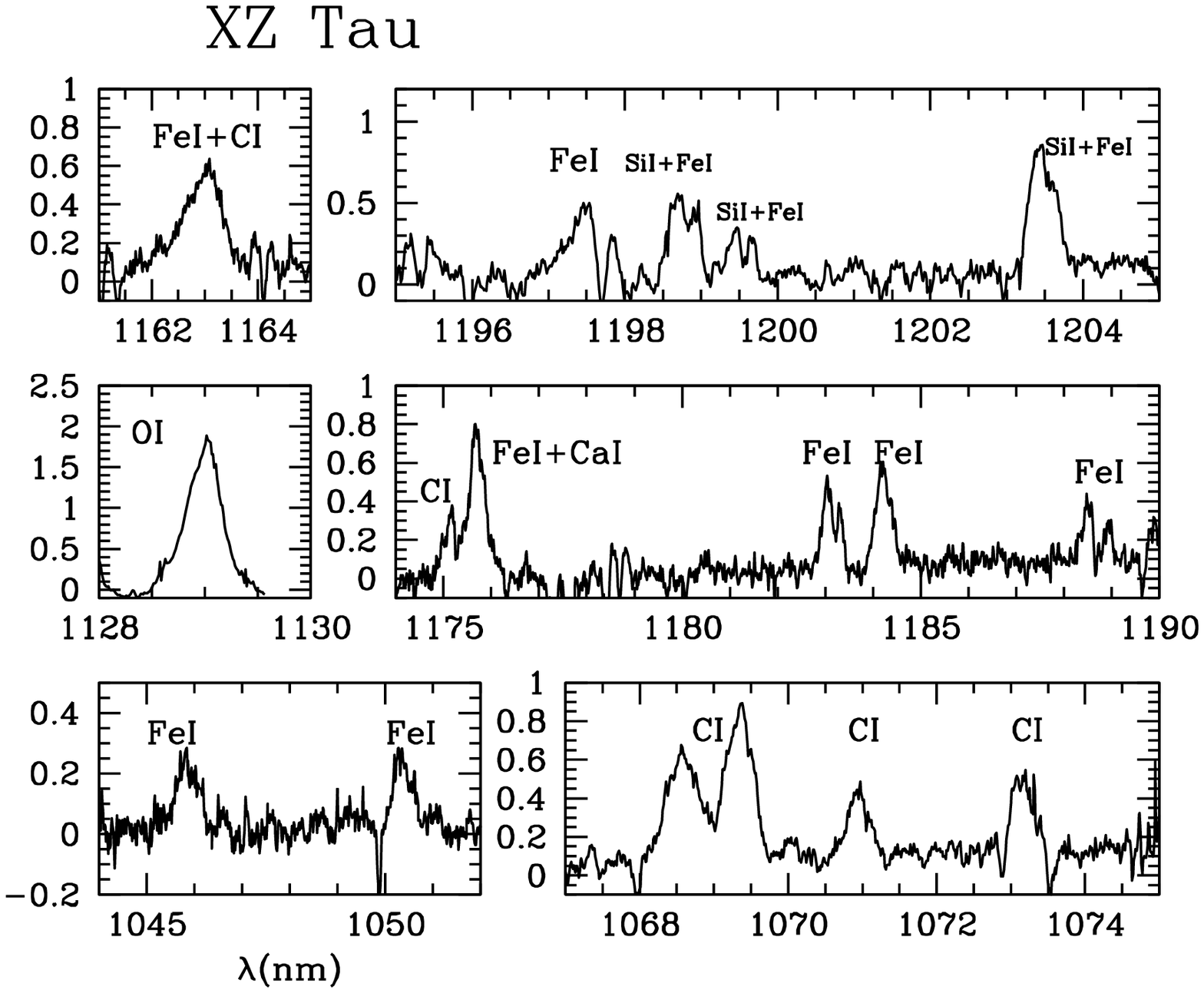}
      \includegraphics[angle=0, width =14.5cm,trim = 0 1.5cm 0 12cm,clip]{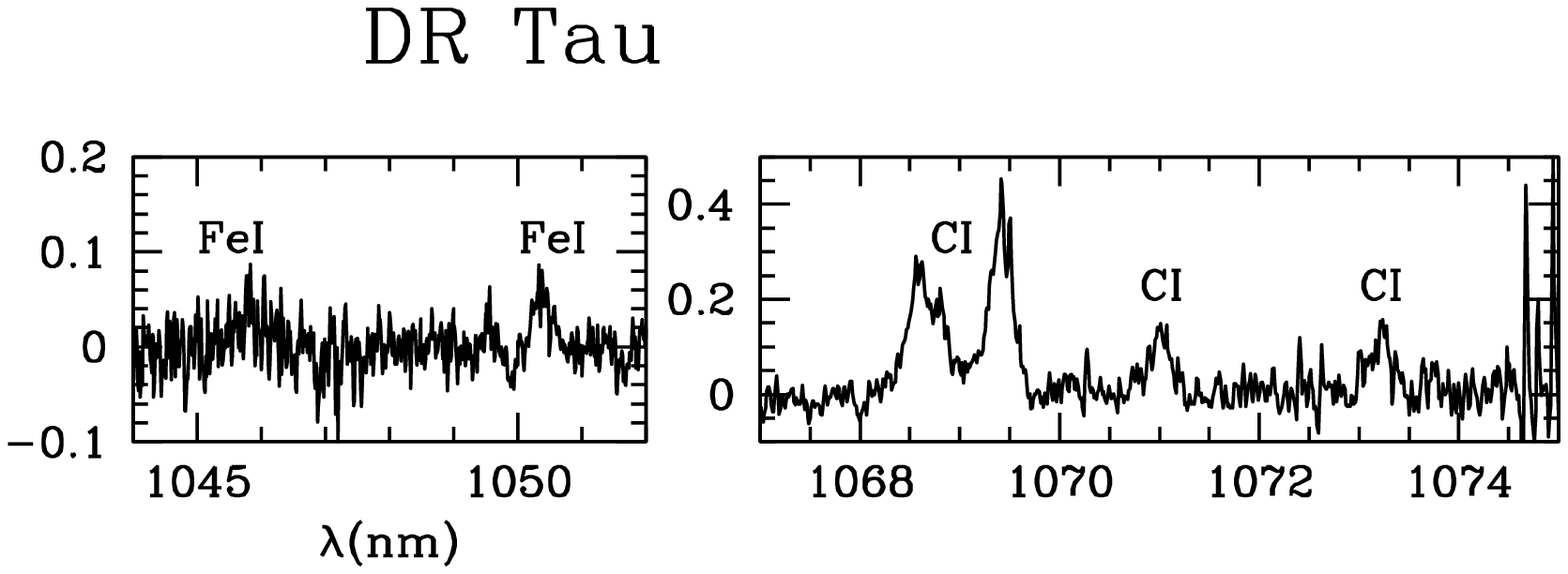}
      \caption{Continuum-subtracted portions of the XZ Tau and DR Tau spectra showing 
      examples of detected metallic lines.
      }
      \label{metallic}
\end{figure*}

\subsection{Metallic lines}

Strong emission is also seen in neutral permitted lines of \fei, \ion{C}{i}, \ion{Si}{i}, \ion{O}{i}, and
\ion{Ca}{i} (see Table 1). They cover excitation energies from $\sim$ 3eV up to $\sim$ 10 eV. 
Neutral species are detected: Given the low ionization potential of some of the
observed species (e.g. Fe: 7.87 eV, and Si: 8.15 eV), this occurrence indicates that 
the emission region of these lines should be predominantly neutral. 

We note that the emission spectra of XZ Tau and
DR Tau are quite different, as in DR Tau we detect a much lower number of metallic lines.
The lines in common between the two sources are mainly the \ion{C}{i} transitions, which are
among the brightest lines in the spectrum.  The EW of these lines is comparable 
in the two objects, while, for example, DR Tau has on average larger EW in the \hi\ lines. 
The width of the DR Tau metallic lines is also systematically smaller than in XZ Tau.
We suggest that the detection of only few metallic lines in DR Tau, as
compared to XZ Tau, is mainly due to an intrinsic difference in the excitation 
conditions of the line emission regions in the two stars.

All \fei\ lines detected in XZ Tau have relatively high upper energy levels (up to 7.45 eV), close to the 
\fei\ ionisation potential (7.87 eV).  This differs from the case, for example, of 
the EXor V1118 Ori (Giannini et al. 2017), whose near-IR spectrum in burst
shows \fei\ lines coming only from lower energy levels, that is, between 3 and 6 eV. 
Also the optical spectrum of DR Tau shows \fei\ lines originating from levels at energies
between 3 and 6eV (Berinstain et al. 1998). We therefore suggest that the non-detection of \fei\
in our DR Tau spectrum could be associated with lower excitation conditions with respect to XZ Tau
and that the lower excitation \fei\ lines, that are detected in V1118 Ori during the burst phase, 
are not seen in DR Tau due to their intrinsic weakness with respect to the optical lines.  
 
Sicilia-Aguilar et al. (2012) present the spectrum of EX Lup for both quiescence and
outburst, showing that during the burst, the number of detected metallic lines tends 
to increase and that the lines broaden. 
The larger number of broader lines detected in XZ Tau at high excitation with respect to DR Tau
could then be related to the particularly active phase of this source, as
we will also discuss in Sect.~\ref{sec:hei}. We however point out that at the date of the
observations XZ Tau had a lower accretion luminosity and veiling than DR Tau, 
therefore it is unlikely that accretion alone is at the origin of the difference in the 
line excitation of the two objects.

\subsection{Line widths}

\begin{figure}[t!]
      \centering
\includegraphics[angle=0, width =8cm, trim = 0cm 1.5cm 0cm 3cm, clip]{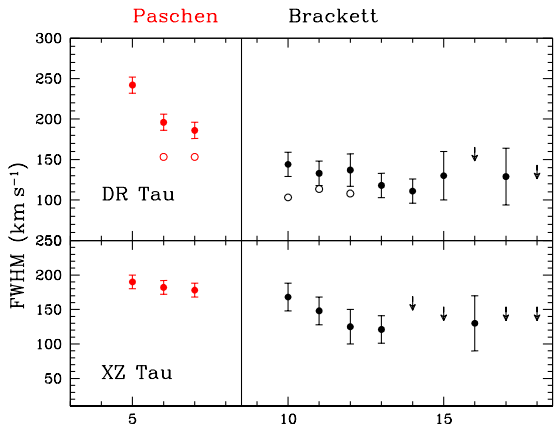}
      \caption{Measured FWHM of \hi\ Paschen and Brackett lines in DR Tau (upper) and XZ Tau (bottom). Upper limits are indicated by arrows. Open dots in the DR Tau plot indicate the width 
      of the main line component once the blue-shifted component has been removed through
      profile decomposition. }
      \label{fig:widths}
\end{figure}

In Fig.~\ref{fig:widths} we plot the width of the HI lines in the two sources, measured as the 
FWHM of the associated Gaussian fit, as a function of the upper level quantum number. 
The figure shows that the line widths tend to decrease with the \textit{n}-value, 
at least up to n$_{up} \sim$ 12. A similar effect was observed also in the EXor variable V1118 Ori,
where Balmer, Paschen, and Brackett lines were considered (Giannini et al. 2017),
and cannot be reconciled with emission from magnetospheric accretion
alone. In fact, in this framework the velocity is increasing as the ionised gas
approaches the stellar surface, thus optically thick lines at high excitation
energy should be characterised by larger widths.

The effect of decreasing width upper quantum number can be in part explained by the presence,
in addition to the accretion component, of a wind component, contributing in
different proportion to the line profiles due to their different decrements
(e.g. Antoniucci et al. 2017). For DR Tau, we additionally plot the FWHM 
of the principal component resulting from the line deconvolution shown in Fig. \ref{brfit}
(open dots). The evidence of decreasing widths, although still present, is now less
prominent. An additional effect could be due to the strong self-absorptions 
characterising the Paschen lines, and the Pa$\beta$ in particular, causing the 
line widths to be overestimated. This is particularly evident in the DR Tau Pa$\beta$ line, 
where the observed double peak is clearly ascribed to self-absorption, as the more
symmetric lines with higher n$_{up}$ are centered around zero velocity, where Pa$\beta$
has a relative minimum.

The widths of the metallic transitions significantly vary among lines, and range from $\sim$40 to 150 \kms. 
The line widths are too large to be compatible with pure chromospheric emission. On the other hand,
gas from the accretion funnel is expected to be ionised and to produce broader lines, 
such as the HI lines. The observed widths of the neutral metallic lines are consistent with 
an origin in the post-shock region of the accretion columns or in the inner 
disk, where the gas is heated to a few thousand K 
(Berinstain et al. 1998; Sicilia-Aguilar et al. 2012; Giannini et al. 2017).

\begin{figure}[t]
      \centering
      \includegraphics[angle=0, width =9.5cm, trim = 0cm 0.5cm 0 1cm,clip]{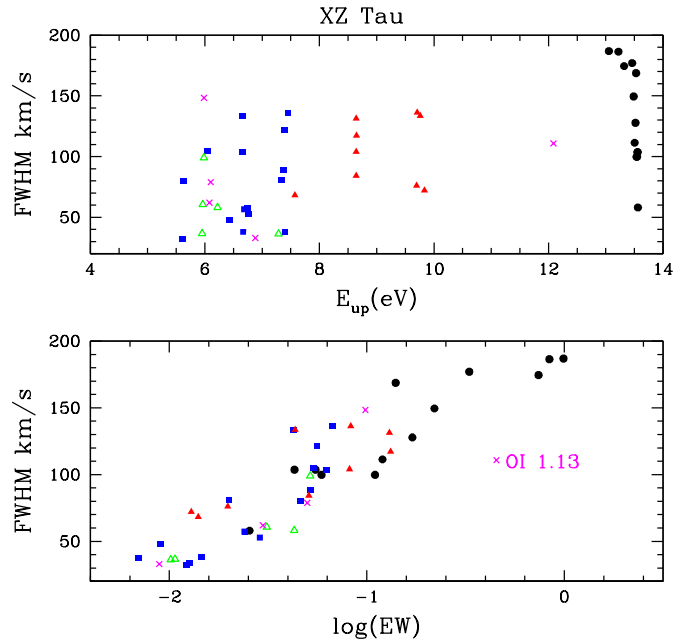}
      \caption{FWHM of the permitted lines detected in XZ Tau plotted
      as a function of the log of the EW (bottom) and the energy
      (in eV) of the upper level (top). Symbols and colour-codes are as follows: black
      filled circles = \hi; blue filled squares = \fei; red filled triangles = \ion{C}{i};
      green open triangles = \ion{Si}{i}; purple crosses = others.
      }
      \label{perm}
\end{figure}

In Fig.~\ref{perm} we report the FWHM of the XZ Tau permitted lines as a function of the 
energy level (upper panel) and line EW (bottom panel).
Line widths do not show any appreciable trend with species or energy level when all
the transitions are considered. 
The plot only very roughly suggests that the narrower lines ($\sim$40-60 \kms) are 
located at energies between 5 and 7 eV, while the broader lines ($\sim$150-200 \kms) are
the \hi\ lines (the Paschen and low-n Brackett in particular). 
The bottom panel of Fig. \ref{perm} shows instead that the FWHM is well 
correlated with the line EW, irrespective of the considered species. A similar 
correlation was previously reported by Beristain et al. (1998) for the 
\fei\ optical lines detected on DR Tau. The authors interpreted this correlation
as due to the presence of two components in the line profiles, a narrow 
and a broad component, that contribute in different proportions among lines with
different intensity. In practice, the broad component contributes to the
overall line broadening only when the signal-to-noise ratio is sufficiently high that 
the line wings are well detected.
In XZ Tau, we see that this correlation extends to a wider range
of species, including the \hi\ lines. Some of the detected lines have profiles that
are not well reproduced by a single Gaussian, such as, for example, the \hi\ lines discussed
in the previous section. However, in most cases it is not possible to
disentangle different components in the line profiles.
We believe that, in addition to the presence of double components, other
causes could be at the origin of the found correlation; namely optical depth effects 
(lines with high optical depth can be auto-absorbed, thus showing larger FWHM)
and the presence of unsorted blending of different lines in broad 
metallic lines.
   
   \begin{figure*}[t]
      \centering
\includegraphics[angle=0, width =9cm, trim = 0cm 0.5cm 8cm 1cm,clip]{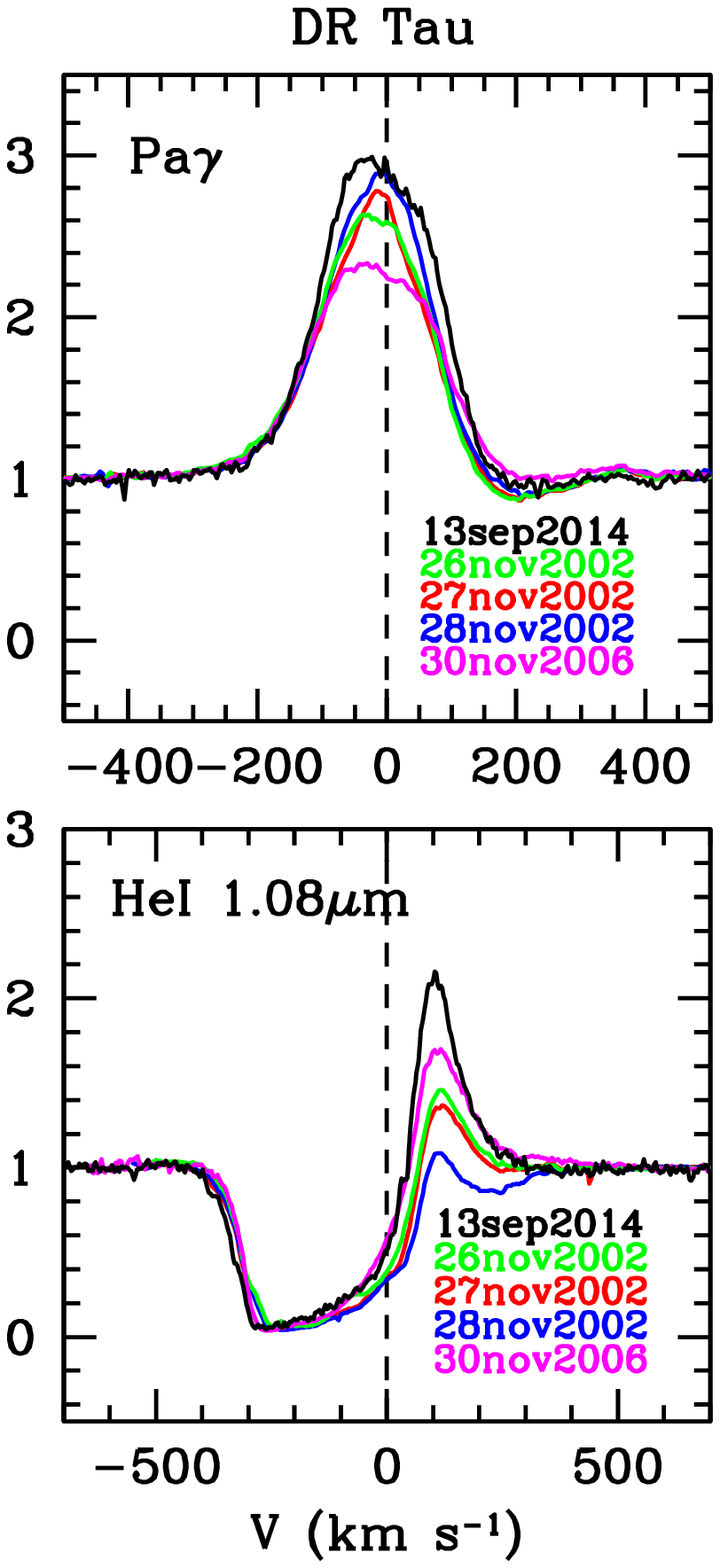}
\includegraphics[angle=0, width =9cm, trim = 0cm 0.5cm 8cm 1cm,clip]{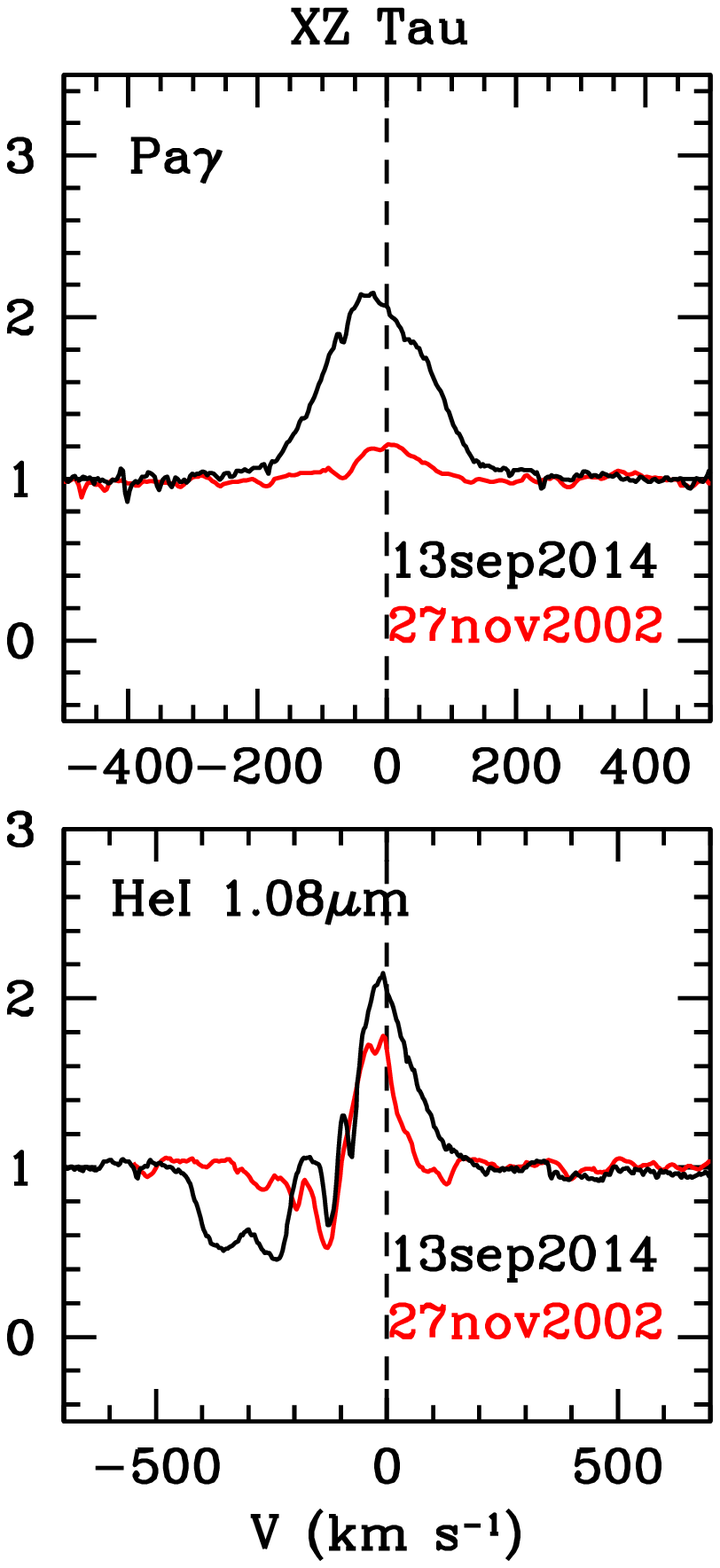}
      \caption{Left: Continuum normalised spectra of the \hei\ 1.08$\mu$m(bottom panel)
      and Pa$\beta$(upper panel) line profiles of DR Tau observed in different epochs. Black
      curves refer to the more recent GIANO spectra presented in this paper, 
      whereas coloured curves are Keck spectra from Edwards et al. (2006)
      acquired on different dates in 2002 and 2006. Right: Same for XZ Tau.}
      
         \label{fig:hei}
   \end{figure*}

\section{The helium 1.08 $\mu$m line: origin and variability}
\label{sec:hei}

Figure~\ref{fig:hei} shows the \hei\ 1.08$\mu$m line in the two sources.
In both cases, the \hei\ line has a composite profile,
with an emission component roughly centred around the systemic velocity,
and a prominent blue-shifted absorption below
the continuum. This last feature is common in the \hei\ profiles of T Tauri
stars and is a recognised signature of hot winds close
to the star (Edwards et al. 2003). The width and depth of this absorption provide direct 
constraints on the inner winds geometry and acceleration region
(Kwan et al. 2007; Kurosawa et al. 2011).
Spectra of the \hei\ line at different epochs are available for both XZ Tau and DR Tau, 
and some insight on the origin of the observed line can be obtained not 
only from its shape but also from its temporal variations.
In Fig.~\ref{fig:hei} in particular, we compare our observed profiles with those obtained by Edwards et al. (2006) with NIRSPEC on Keck II ($R$=25,000). These figures also show the variations of the Pa$\gamma$ profile in the same epochs. 

For XZ Tau, our spectrum is compared with the Keck spectrum acquired in Nov. 2002. 
The main feature of the profile observed in Nov. 2002 was that the blue-shifted absorption was not continuous 
but appeared to be composed by several absorption features with different depths at different velocities,
up to 300 km/s. This kind of profile can be interpreted as
originating in a disk-wind. Indeed, in contrast to a radially symmetric expanding stellar wind, 
in a disk wind the continuum from the star will intercept a much narrower
range of velocities, which will be confined to nearly parallel
streamlines emerging from the disk surface. The fact that different narrow
absorptions are present in the XZ Tau profile indicates that the disk wind has a complex 
and non-uniform structure. With respect to 2002, the \hei\ profile observed by GIANO
in 2014 shows a much broader and deeper blue-shifted absorption, extending up to 500 km/s.
In this time span the \hei\ emission component has also broadened and brightened with a
simultaneous increase of the Pa$\gamma$ equivalent width by a factor
more than 10, which passed from EW = 1 \AA\ (Edwards et al. 2003) to EW = 16 \AA\ in 2014.
These evidences, together with the increases in the 1\um\, veiling already discussed 
in Sect.~\ref{sec:veiling_rotvel}, hint at line profile changes that are connected with mass accretion
rate variability: an increase in the mass accretion rate results in brightening 
of accretion-related emission lines, such as the Pa$\gamma$, and of the 1\um\, excess 
causing the line veiling. 
At the same time, accretion might have induced a dramatic
change in the wind structure, and/or the onset of a new flow component very
close to the star (i.e. at a speed close to the stellar escape velocity).
As already discussed in Sect.~\ref{sec:lines}, the fact that XZ Tau was in 
a particularly high state of accretion is also testified by the higher 
mass-accretion rate derived from our data when compared to that implied 
by previous Pa$\beta$ observations (Lorenzetti et al. 2009). In addition,
the $H$-band magnitude at the time of our observations was significantly 
lower than that measured in November 2002 (i.e. 7.6 against 8.3 mag, Hioki et al. 2009).
We cannot discern, from these two epoch observations, the timescales for the observed
variability. Chou et al. (2013) report short term variability
of emission line profiles on daily and monthly bases, in addition to a long-term 
variability on 3-10 yr timescales. Additional observations are needed to understand
with which of the above variability scales the observed wind variability is associated.

In the case of DR Tau, several comparison spectra are available, acquired both in 2002 and in 2006.
The profile of the \hei\ line presents significant differences with respect to that of XZ Tau. 
DR Tau shows indeed a broad and deep blue-shifted absorption extending to -350 \kms, which
suggests the formation in a stellar wind. In fact, the broad range of 
velocities covered by the absorption feature can be explained by a radially emerging wind absorbing
the 1 \um\, continuum from the stellar surface, thus tracing the full
acceleration region of the inner wind (Kwan et al. 2007).
Noticeably, the blue-shifted absorption component does not significantly change in the various spectra.
At variance, the line component in emission is subject to 
significant variations on time-scales of days, indicating that the emission
component mainly arises from a mechanism different from the wind itself.
As also discussed in Sect.~\ref{sec:lines}, variability of \hi\ lines have been observed in DR Tau, 
with periods ranging from 4 to 9 days (Alencar et al. 2001). 
Such a variability has been linked to shock emission in hot spots combined with stellar rotation. 
The emission component of the \hei\ line could be imputed to the same mechanism, 
although the lack of correlation between the variation of the \hei\ emission 
and that of Pa$\gamma$ likely indicates that the two lines react in different ways to 
changes in the accretion shock conditions. Balmer line profiles (Alencar et al. 2001) show extended 
blue-shifted wing components with a maximum velocity of more than 400\kms . However
and at variance with the \hei\ 1.08\um\ line, these components show a strong 
variability as their intensity
and peak velocity change on timescales of a few days, passing from absorption to emission. 
It is therefore unlikely that they are related to the very stable stellar
wind responsible for the deep absorption seen in the \hei\ line.
We finally note that when the \hei\ emission component is significantly fading, 
a red-shifted absorption is observed, peaking at roughly the same velocity ($\sim$ 200\kms)
as the inverse-P Cygni observed in the Pa$\gamma$ profile. This absorption feature is
probably cancelled out when the broad emission component becomes stronger.

The reason why the sub-continuum absorption of the two sources has such a different
behaviour (i.e. a steady stellar wind in DR Tau and a highly variable 
disk wind in XZ Tau) remains unclear. Geometry may play a role in this respect. 
DR Tau is known to be close to pole-on, meaning that we are probably looking at 
the polar region of a stellar wind (either spherical or conical). On the other 
hand, XZ Tau is inclined by $\sim$ 30$\degr$ with respect to the plane of the sky, 
consequently the disk wind is more directly viewed. Disk winds seem
therefore to react to accretion variability more directly than stellar winds.

\begin{figure}[t]
   \centering
   \includegraphics[angle=0, width =15cm, trim = 0.5cm 1.5cm 0 1cm,clip]{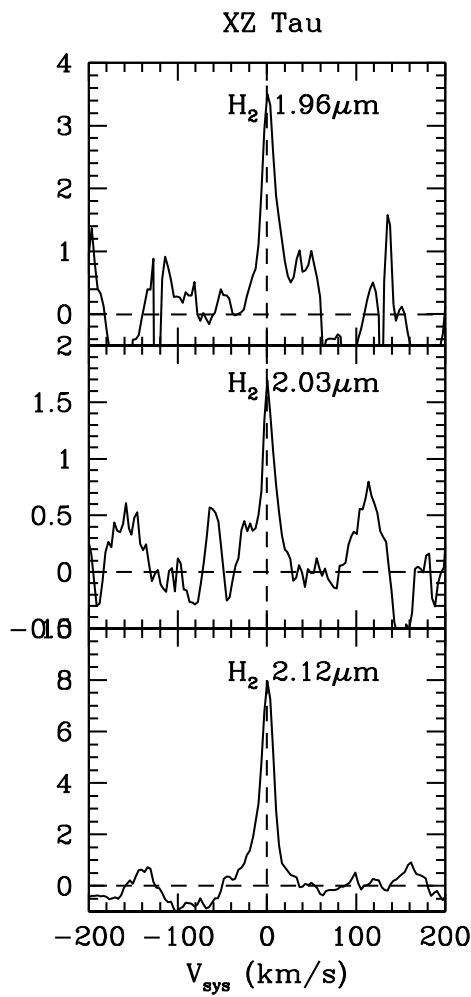}
   \caption{Observed H$_2$ lines in XZ Tau.
   }
   \label{fig:h2}
\end{figure}

\begin{figure}[t]
      \centering
      \includegraphics[angle=0, width =14cm,trim = 1cm 1.5cm 0 10cm,clip]{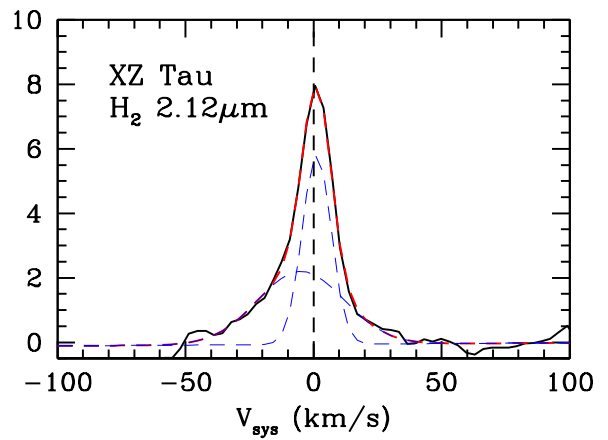}
      \caption{Profile of the H$_2$ 2.12\um\, line detected in XZ Tau with the 
      fit of the two Gaussian components
superimposed. The red-dashed line is the best fit obtained by the sum of 
      the two Gaussian components plotted as blue-dashed lines. Parameters of the two components
      are given in Table 2.
      }
      \label{fig:h2_deconv}
\end{figure}   

\section{H$_2$ lines in XZ Tau}
\label{sec:h2}

The GIANO spectrum of XZ Tau shows emission from three H$_2$ v=1-0 
ro-vibrational lines (Fig.~\ref{fig:h2}). The lines are narrow but resolved
at the 6 \kms\ resolution of GIANO. The 2.12 \um\, line, in particular, 
shows a profile that is not Gaussian but presents prominent wings.
We deconvolved this profile into two Gaussian components, a narrow one, centred
close to systemic velocity, and a blue-shifted and broader component (see Fig.~\ref{fig:h2_deconv})
The parameters of this deconvolution are given in Table 2. The narrow 
component has a width of $\sim$ 13 \kms, which, taking into account the
instrument resolution, implies an intrinsic width of 11 \kms. This small
line width and the fact that the line does not show any appreciable velocity
shift with respect to quiescence suggest that this emission might originate
in a disk and be broadened as a result of the Keplerian rotation
of the gas. 
Similar quiescent H$_2$ emission has been observed in other
low-mass T Tauri stars, for which an origin from excitation of the disk gas by UV or X-rays stellar photons was suggested 
(e.g. Bary et al. 2003, 2008).
ALMA 1.3mm observations of the XZ Tau system have resolved a compact disk of radius 3.2 AU
 around XZ Tau B (Osorio et al. 2016). In the hypothesis that the quiescent H$_2$ emission
we observe originates from this disk, we can estimate the innermost H$_2$ emission
radius from the measured profile, that is, $R_{in} = GM_*\times (sin(i)/\Delta V)^2$.
Here $i$ is the disk inclination angle, which we take to be equal to 35$\degr$ (Osorio et al. 2016),
$M_* = 0.37$ M$_{\odot}$, and $\Delta V$ is the maximal velocity of the $H_2$
emitting gas, which we take to be equal to 1.7$\times$HWHM following the prescription by Salyk et al. (2011).
We derive $R_{in} \sim$ 1.2 AU, which is about equal to the inner radius of the 1.3mm disk
observed by ALMA (1.3 AU, Osorio et al. 2016). Hence in this hypothesis, the near-IR H$_2$
emission is excited in the inner rim of the dusty disk. 
If, instead, part of the narrow emission comes also from the A source, the derived
inner radius can be considered as a lower limit and one should conclude that 
the H$_2$ originates from a region of the disk where both gas and dust coexist.

The blue-shifted broad component has instead a line width of about 39\kms, which
is typical of shock-excited gas (Flower et al. 2003).
Beck et al. (2008), present images of the H$_2$ spatial distribution in the surrounding
of XZ Tau showing extended line emission both around the sources and
in an arc-shaped structure located at about 0\farcs 5 from the binary. We suggest that the broad component
originates from this structure, which likely represents gas shock-excited in the XZ Tau
expanding wide-angle flow. We also remark that the radial velocity of this blue-shifted 
component is compatible with the velocity of the CO(1-0) emission associated with
the outflowing gas as observed by ALMA (Zapata et al. 2015).

\begin{table}[t]  
\caption{Parameters of the H$_2$ lines}
\label{tab:h2}
\begin{center}  
\begin{tabular}{cccc}
\hline
\hline
\noalign{\smallskip}
  $\lambda$ & ID & V$_{sys}$ & FWHM  \\
   (nm) & & (\kms) & (\kms) \\ 
\noalign{\smallskip}
\hline
\noalign{\smallskip}
1957.56$^a$ & 1-0S(3) & 2.0  & 19.8 \\
2033.76$^b$ & 1-0S(2) & 1.9  & 12.5 \\
          &           &$-$6.7  & 38.5 \\
2121.83$^b$ & 1-0S(1) & 1.2  & 13.2 \\
          &           &$-$5.0  & 39.8 \\
\noalign{\smallskip}
\hline
\end{tabular}
\tablefoot{
\tablefoottext{a}{Single Gaussian component fit.}
\tablefoottext{b}{Two Gaussian components fit.}
}
\end{center}
\end{table}

\section{Conclusions}
\label{sec:conclusions}

We have presented near-IR (0.95-2.4\um) high-resolution ($R$ = 50\,000) spectra of the
two EXor-type variables XZ Tau and DR Tau, obtained with the GIANO instrument on TNG. 
The analysis of the emission and absorption features in these spectra 
allowed us to investigate the gas properties in the interaction region between the
star and the inner accretion disk.

Different evidences show that XZ Tau was in a fairly high state of accretion during the GIANO
observations. This is suggested in particular by a comparison 
with Keck IR spectra acquired in the past. 
With respect to previous observations, we measure an increase of both IR veiling 
and EW of the Pa$\gamma$ line. Simultaneous variations of the blue-shifted 
absorption feature in the \hei\ 1.08\um\ line suggest that the inner wind 
has undergone a dramatic change in its velocity structure. From the Pa$\beta$
luminosity, calibrated with almost simultaneous photometry, we estimate a
mass accretion rate \macc = 4.2$\times$10$^{-8}$ \msunyr. 
DR Tau shows a lower than normal \lacc, although still higher than the one displayed by XZ Tau in its enhanced accretion state. 
Both sources are in any case CTT stars that display high accretion luminosity values ($L_{acc}$ = 0.34 for XZ Tau 
and 0.79 \lsun for DR Tau).

\hi\ and \hei\ lines are not symmetric in both sources, thus testifying
the presence of different components (from accretion and wind regions) contributing to the line profiles.
Paschen lines are broad ($\Delta$V$_{sys}$ $\sim$ 180 and 200 \kms\, in XZ Tau and DR Tau, respectively) 
and show red-shifted asymmetries in both sources. Inverse P-Cygni
absorption is observed in the Pa$\gamma$ of DR Tau. Brackett lines are more symmetric and
narrower. In DR Tau, two emission components are clearly recognisable in Paschen and Brackett lines of higher SNR,
one centred at velocity close to zero and the other blue-shifted at V$_{sys}=-$100 \kms.
This latter can be associated with emission from a high-velocity wind 
previously underlined in the profiles of Balmer lines (Alencar et al. 2001). 
A prominent high-velocity stellar wind is also traced by the broad and deep
absorption feature of the \hei\ 1.08 \um\, line. At variance with XZ Tau, 
this feature appears stable in time, in spite of significant variations 
not only of the emission component of the \hei\ and Pa$\gamma$ lines, but also 
of the blue-shifted wind component probed by the Balmer lines. This suggests that
the inner hot region of the stellar wind traced by the \hei\ absorption 
is not affected by short-term variations in the accretion and disk-wind
flows occurring in different regions and on different spatial scales.  


From the comparison of the target photospheric spectra with that of 
template spectra of comparable spectral types, we were able to measure the IR
veiling as a function of wavelength. The spectral energy distribution 
of the IR veiling can be fitted with a single blackbody having  
emitting areas $\sim$ 30 times the stellar surface. 
In XZ Tau the fitted blackbody has a temperature T=1600 K.
The 1\um\, veiling is higher than
measured by Edwards et al. (2002), but the shape and blackbody temperature are
similar to those found in other T Tauri stars of similar luminosity.
The finding is consistent with emission from the inner region of a circumstellar 
disk where the dust starts to sublimate. DR Tau shows a much larger veiling, as
previously reported in the literature, consistent with a blackbody at T=2300 K.
At this temperature we expect the dust to be already destroyed, therefore
the DR Tau IR excess is more consistent with emission from an optically thick 
gaseous disk. 
The possibility that part of or all the IR veiling in the
range of temperatures between 2000 and 5000 K might arise from 
an inner dust-free gaseous disk has been discussed by Fisher et al. (2011),
who recognised that CTT stars often show an excess emission in the \textit{IYJ} bands
consistent with this range of temperatures.

In the XZ Tau spectrum we detect strong permitted metallic lines from neutral species
such as \fei, \ion{C}{i}, \ion{Ca}{i}, and \ion{Si}{i}, whereas the DR Tau spectrum shows only a few
narrower metallic lines, mainly from \ion{C}{i}. No ionised species are observed
in either source. In general, the metallic lines in 
XZ Tau seem to originate in a gas that is more excited than in DR Tau. 
This evidence excludes that the excitation of these lines is mainly driven by accretion,
as DR Tau has a larger veiling and accretion luminosity than XZ Tau. 

The \htwo\ 2.12 \um\, line observed in XZ Tau shows two emission components: A narrow one
centred at zero velocity, which is compatible with an origin from a gaseous
disk, and a blue-shifted broad component that is probably shock-excited in 
the XZ Tau expanding wide-angle flow.

The presented observations demonstrate the potential of wide-band, high-resolution 
near-IR spectroscopy to probe the different phenomena occurring in the region of interaction
between the star magnetosphere and the accretion disk.  
In particular, these observations demonstrate that 
a resolution of 50\,000 is required to kinematically separate
winds and accretion regions in the profiles of permitted lines, or disk-bound gas from shock 
emission in H$_2$ lines. In addition, simultaneous observations at high resolution of the 
entire NIR range open the possibility to study in detail the variations of the IR veiling with wavelength,
hence helping to constrain the properties of the inner disk. The planned upcoming combination of an upgrade 
of GIANO with the HARPS-N optical instrument (the GIARPS facility at TNG, Claudi et al. 2016)
will help to further improve the unified picture of the accretion and ejection phenomena in young disks,
through observation of complementary optical tracers (e.g. veiling from accretion shocks and wind forbidden lines).


\begin{acknowledgements}
We thank the anonymous referee for her/his very helpful suggestions that helped us to greatly improve our paper.\\

We are very thankful to the staff of the Campo Imperatore Observatory for carrying out the photometric observations for the EXORCISM monitoring program.\\

We are grateful to Suzan Edwards for providing the Keck spectra of XZ Tau and DR Tau.\\

KB thanks the Osservatorio Astronomico di Roma for the support given during her visits. KB was financially supported by the PRIN INAF 2013 "Disks, jets, and the dawn of planets".\\

We acknowledge with thanks the variable star observations from the AAVSO International Database contributed by observers worldwide and used in this research.
\end{acknowledgements}

%

\begin{thebibliography}{}

\bibitem[Alcal{\'a} et al.(2014)]{2014A&A...561A...2A} Alcal{\'a}, J.~M., Natta, A., Manara, C.~F., et al.\ 2014, \aap, 561, A2 

\bibitem[Alcal{\'a} et al.(2017)]{2017A&A...600A..20A} Alcal{\'a}, J.~M., Manara, C.~F., Natta, A., et al.\ 2017, \aap, 600, A20 

\bibitem[Alencar et al.(2001)]{2001AJ....122.3335A} Alencar, S.~H.~P., 
Johns-Krull, C.~M., \& Basri, G.\ 2001, \aj, 122, 3335 

\bibitem[Alexander et al.(2014)]{alexander2014} Alexander, R., Pascucci, I., Andrews, S., Armitage, P., and Cieza, L.  The Dispersal of Protoplanetary Disks  2014,  Protostars and Planets VI,   475  
http://adsabs.harvard.edu/abs/2014prpl.conf..475A  

\bibitem[Antoniucci et al.(2013)]{2013prpl.conf2B055A} Antoniucci, S., Arkharov, A.~A., Di Paola, A., et al.\ 2013, Protostars and Planets VI Posters, 2 

\bibitem[Antoniucci et al.(2014)]{2014A&A...572A..62A} Antoniucci, S., Garc{\'{\i}}a L{\'o}pez, R., Nisini, B., et al.\ 2014, \aap, 572, A62 

\bibitem[Antoniucci et al.(2014)]{2014A&A...565L...7A} Antoniucci, S., Arkharov, A.~A., Di Paola, A., et al.\ 2014, \aap, 565, L7 

\bibitem[Antoniucci et al.(2017)]{2017A&A...599A.105A} Antoniucci, S., Nisini, B., Giannini, T., et al.\ 2017, \aap, 599, A105 

\bibitem[Audard et al.(2014)]{2014prpl.conf..387A} Audard, M., {\'A}brah{\'a}m, P., Dunham, M.~M., et al.\ 2014, Protostars and Planets VI, 387 

\bibitem[Banzatti et al.(2014)]{2014ApJ...780...26B} Banzatti, A., Meyer, M.~R., Manara, C.~F., Pontoppidan, K.~M., \& Testi, L.\ 2014, \apj, 780, 26 

\bibitem[Bary et al.(2003)]{2003ApJ...586.1136B} Bary, J.~S., Weintraub, D.~A., \& Kastner, J.~H.\ 2003, \apj, 586, 1136 

\bibitem[Bary et al.(2008)]{2008ApJ...678.1088B} Bary, J.~S., Weintraub, D.~A., Shukla, S.~J., Leisenring, J.~M., \& Kastner, J.~H.\ 2008, \apj, 678, 1088-1098 

\bibitem[Beck et al.(2008)]{2008ApJ...676..472B} Beck, T.~L., McGregor, P.~J., Takami, M., \& Pyo, T.-S.\ 2008, \apj, 676, 472-489 

\bibitem[Beristain et al.(1998)]{1998ApJ...499..828B} Beristain, G., Edwards, S., \&  Kwan, J.\ 1998, \apj, 499, 828

\bibitem[Caffau et al.(2016)]{2016A&A...585A..16C} Caffau, E., Andrievsky, S., Korotin, S., et al.\ 2016, \aap, 585, A16 

\bibitem[Calvet \& Gullbring(1998)]{1998ApJ...509..802C} Calvet, N., \& Gullbring, E.\ 1998, \apj, 509, 802 

\bibitem[Carleo et al.(2016)]{2016ExA....41..351C} Carleo, I., Sanna, N., Gratton, R., et al.\ 2016, Experimental Astronomy, 41, 351 

\bibitem[Carrasco-Gonz{\'a}lez et al.(2009)]{2009ApJ...693L..86C} Carrasco-Gonz{\'a}lez, C., Rodr{\'{\i}}guez, L.~F., Anglada, G., \& Curiel, S.\ 2009, \apjl, 693, L86 

\bibitem[Chou et al.(2013)]{2013AJ....145..108C} Chou, M.-Y., Takami, M., 
Manset, N., et al.\ 2013, \aj, 145, 108 

\bibitem[Claudi et al.(2016)]{2016SPIE.9908E..1AC} Claudi, R., Benatti, S., Carleo, I., et al.\ 2016, \procspie, 9908, 99081A 

\bibitem[Claret \& Bloemen(2001)]{claretbloemen2011} Claret, A., \& Bloemen, S. 2011, \aap, 529, A75

\bibitem[D'Alessio et al.(2000)]{2000SPIE.4008..748D} D'Alessio, F., Di Cianno, A., Di Paola, A., et al.\ 2000, \procspie, 4008, 748 

\bibitem[Edwards et al.(2006)]{2006ApJ...646..319E} Edwards, S., Fischer, 
W., Hillenbrand, L., \& Kwan, J.\ 2006, \apj, 646, 319 

\bibitem[Edwards et al.(2003)]{2003ApJ...599L..41E} Edwards, S., Fischer, W., Kwan, J., Hillenbrand, L., \& Dupree, A.~K.\ 2003, \apjl, 599, L41 

\bibitem[Eisner et al.(2005)]{2005ApJ...623..952E} Eisner, J.~A., Hillenbrand, L.~A., White, R.~J., Akeson, R.~L., \& Sargent, A.~I.\ 2005, \apj, 623, 952 

\bibitem[Fischer et al.(2011)]{2011ApJ...730...73F} Fischer, W., Edwards, S., Hillenbrand, L., \& Kwan, J.\ 2011, \apj, 730, 73 

\bibitem[Flower et al.(2003)]{2003MNRAS.341...70F} Flower, D.~R., Le Bourlot, J., Pineau des For{\^e}ts, G., \& Cabrit, S.\ 2003, \mnras, 341, 70 

\bibitem[Folha \& Emerson(1999)]{1999A&A...352..517F} Folha, D.~F.~M., \& Emerson, J.~P.\ 1999, \aap, 352, 517 

\bibitem[Folha \& Emerson(2001)]{2001A&A...365...90F} Folha, D.~F.~M., \& Emerson, J.~P.\ 2001, \aap, 365, 90 

\bibitem[Frasca et al.(2015)]{frascaetal2015} Frasca, A., Biazzo, K., Lanzafame, A. C., et al. 2015, \aap, 575, A4

\bibitem[Giannini et al.(2017)]{2017ApJ...839..112G} Giannini, T., Antoniucci, S., Lorenzetti, D., et al.\ 2017, \apj, 839, 112 

\bibitem[Gray(2005)]{Gray2005} Gray, D. F. 2005, {\it The Observation and Analysis of Stellar Photospheres}, 3$^{\rm rd}$ ed., Cambridge University Press

\bibitem[Hartigan \& Kenyon(2003)]{hartigankenyon2003} Hartigan, P., \& Kenyon, S. J., 2003, \apj, 583, 334

\bibitem[Johns-Krull \& Valenti(2001)]{2001ASPC..244..147J} Johns-Krull, C.~M., \& Valenti, J.~A.\ 2001, Young Stars Near Earth: Progress and Prospects, 244, 147 

\bibitem[Kurosawa et al.(2011)]{2011MNRAS.416.2623K} Kurosawa, R., 
Romanova, M.~M., \& Harries, T.~J.\ 2011, \mnras, 416, 2623 

\bibitem[Kwan et al.(2007)]{2007ApJ...657..897K} Kwan, J., Edwards, S., 
\& Fischer, W.\ 2007, \apj, 657, 897 

\bibitem[Lorenzetti et al.(2012)]{2012ApJ...749..188L} Lorenzetti, D., Antoniucci, S., Giannini, T., et al.\ 2012, \apj, 749, 188 

\bibitem[Lorenzetti et al.(2009)]{2009ApJ...693.1056L} Lorenzetti, D., Larionov, V.~M., Giannini, T., et al.\ 2009, \apj, 693, 1056 


\bibitem[Manara et al.(2017)]{2017arXiv170402842M} Manara, C.~F., Testi, L., Herczeg, G.~J., et al.\ 2017, arXiv:1704.02842 


\bibitem[McClure et al.(2013)]{2013ApJ...769...73M} McClure, M.~K., Calvet, N., Espaillat, C., et al.\ 2013, \apj, 769, 73 

\bibitem[Muzerolle et al.(2001)]{2001ApJ...550..944M} Muzerolle, J., Calvet, N., \& Hartmann, L.\ 2001, \apj, 550, 944 

\bibitem[Muzerolle et al.(2003)]{2003ApJ...597L.149M} Muzerolle, J., Calvet, N., Hartmann, L., \& D'Alessio, P.\ 2003, \apjl, 597, L149 

\bibitem[Nguyen et al.(2012)]{nguyenetal2012} Nguyen, D. C., Brandeker, A., van Kerkwijk, M. H., \& Jayawardhana, R. 2012, \apj, 745, 119

\bibitem[Nisini et al.(2004)]{2004A&A...421..187N} Nisini, B., Antoniucci, S., \& Giannini, T.\ 2004, \aap, 421, 187 

\bibitem[Nisini et al.(2005)]{2005A&A...441..159N} Nisini, B., Bacciotti, F., Giannini, T., et al.\ 2005, \aap, 441, 159 

\bibitem[Oliva et al.(2012)]{2012SPIE.8446E..3TO} Oliva, E., Origlia, L., Maiolino, R., et al.\ 2012, \procspie, 8446, 84463T 

\bibitem[Origlia et al.(2014)]{2014SPIE.9147E..1EO} Origlia, L., Oliva, E., Baffa, C., et al.\ 2014, \procspie, 9147, 91471E 

\bibitem[Osorio et al.(2016)]{2016ApJ...825L..10O} Osorio, M., Mac{\'{\i}}as, E., Anglada, G., et al.\ 2016, \apjl, 825, L10 

\bibitem[Petrov et al.(2011)]{2011A&A...535A...6P} Petrov, P.~P., Gahm, G.~F., Stempels, H.~C., Walter, F.~M., \& Artemenko, S.~A.\ 2011, \aap, 535, A6 

\bibitem[Rayner et al.(2009)]{rayneretal2009} Rayner, J. T., Cushing, M. C., \& Vacca, W. D. 2009, \apjs, 185, 289

\bibitem[Rigliaco et al.(2013)]{2013ApJ...772...60R} Rigliaco, E., Pascucci, I., Gorti, U., Edwards, S., \& Hollenbach, D.\ 2013, \apj, 772, 60 

\bibitem[Salyk et al.(2011)]{2011ApJ...743..112S} Salyk, C., Blake, G.~A., Boogert, A.~C.~A., \& Brown, J.~M.\ 2011, \apj, 743, 112 

\bibitem[Sharon et al.(2010)]{sharonetal2010} Sharon, C., Hillenbrand, L., Fischer, W., \& Edwards, S. 2010, \apj, 139, 646

\bibitem[Sicilia-Aguilar et al.(2012)]{2012A&A...544A..93S} Sicilia-Aguilar, A., K{\'o}sp{\'a}l, {\'A}., Setiawan, J., et al.\ 2012, \aap, 544, A93 

\bibitem[Tonry \& Davis(1979)]{tonrydavis1979} Tonry, J., \& Davis, M. 1979, \apj, 84, 1511

\bibitem[Tozzi et al.(2014)]{2014SPIE.9147E..9NT} Tozzi, A., Oliva, E., Origlia, L., et al.\ 2014, \procspie, 9147, 91479N 

\bibitem[White \& Hillenbrand(2004)]{whitehillenbrand2004} White, R. J., \& Hillenbrand, L. A. 2004, \apj, 616, 998

\bibitem[Zapata et al.(2015)]{2015ApJ...811L...4Z} Zapata, L.~A., Galv{\'a}n-Madrid, R., Carrasco-Gonz{\'a}lez, C., et al.\ 2015, \apjl, 811, L4 





\end{thebibliography}
%

\end{document}